\documentclass[12pt]{article}
\usepackage{amsmath}
\usepackage{graphicx}
\usepackage{enumerate}
\usepackage{url} 

\usepackage{amsthm}
\usepackage[authoryear]{natbib}

\usepackage[nottoc]{tocbibind}

\usepackage[colorlinks =true,linkcolor=blue,urlcolor=blue,citecolor=blue]{hyperref}

\usepackage{ragged2e}

\usepackage{subcaption}
\usepackage{url}
\usepackage[utf8]{inputenc}
\usepackage{tabulary}
\usepackage{geometry}
\usepackage{setspace}
\usepackage{authblk}
\usepackage{natbib}
\usepackage{hyperref}
\usepackage{float}
\usepackage{multirow}
\usepackage{graphicx}
\usepackage{adjustbox}
\usepackage{caption}
\usepackage{booktabs}
\usepackage{tabularx}  
\usepackage{amsfonts}
\usepackage{amssymb}
\usepackage{amsxtra}
\usepackage{amsthm}
\usepackage{mathrsfs}
\usepackage{bbm}
\usepackage{enumerate}
\usepackage{enumitem}
\usepackage{xcolor}
\usepackage{lscape}
\usepackage{pgfplots}

\theoremstyle{definition}

\newcommand{\bitemsize}{\begin{itemize}}
\newcommand{\eitemsize}{\end{itemize}}
\newcommand{\be}{\begin{equation*}\begin{aligned}}  
\newcommand{\ee}{\end{aligned}\end{equation*}}
\newcommand{\benum}{\begin{enumerate}}
\newcommand{\eenum}{\end{enumerate}}
\newcommand{\bmat}{\begin{bmatrix}}
\newcommand{\emat}{\end{bmatrix}}

\title{The (Short-Term) Effects of Large Language Models on Unemployment and Earnings}
\author{Danqing Chen, Carina Kane, Austin Kozlowski, Nadav Kuniesvky, and James A. Evans\footnote{University of Chicago, Knowledge Lab. }}

\begin{document}
\maketitle
\begin{abstract}
Large Language Models (LLMs) have spread rapidly since the release of ChatGPT in late 2022, accompanied by claims of major productivity gains but also concerns about job displacement. This paper examines the short-run labor market effects of LLM adoption by comparing earnings and unemployment across occupations with differing levels of exposure to these technologies. Using a Synthetic Difference-in-Differences approach, we estimate the impact of LLM exposure on wages and unemployment. Our findings show that workers in highly exposed occupations experienced earnings increases following ChatGPT’s introduction, while unemployment rates remained unchanged. These results suggest that initial labor market adjustments to LLMs operate primarily through wages rather than worker reallocation.
\end{abstract}

\section{Introduction}

Large Language Models (LLMs) have emerged as one of the most transformative recent technological innovations, fundamentally altering how individuals and organizations approach information processing, content creation, and problem-solving tasks. Since the public release of ChatGPT in November 2022, the adoption of these tools has been unprecedented in both scope and speed, with millions of users integrating LLMs into their daily workflows across diverse sectors of the economy. Despite this rapid proliferation and widespread usage, the economic implications of LLMs, particularly their effects on workers and labor market outcomes, remain poorly understood.

Predicting the labor-market effects of LLMs is difficult because the same technology can plausibly both augment and automate work. In canonical labor-demand terms, the sign of the impact hinges on whether LLMs operate primarily as complements or as substitutes for human skills. When they complement workers by raising task productivity, the labor-demand curve shifts outward, increasing wages and—after market adjustment—employment. When they substitute for workers by automating tasks previously done by humans, the demand curve shifts inward, reducing wages and employment in the affected occupations.

In this paper, we study the impact of LLMs on key labor market outcomes, leveraging the quasi-experimental variation generated by the sudden public release of ChatGPT. Our empirical strategy employs the Synthetic Difference-in-Differences (SDiD) methodology (\cite{arkhangelsky2021synthdid}), which allows us to estimate counterfactual outcomes for occupations highly exposed to LLMs by constructing synthetic control groups from less exposed occupations. To measure occupational exposure, we build on \cite{handa2025economic}, who identify tasks in the Occupational Information Network (O\*NET) that are potentially affected by LLMs. We then assess whether these tasks can be effectively targeted through LLM prompts and define an occupation’s exposure as the share of its tasks that are susceptible to LLM influence.

We find that the introduction of ChatGPT increased earnings for workers in occupations with high exposure to LLMs. However, we do not observe corresponding effects on unemployment in these same occupations. Specifically, we estimate that occupations who are more exposed to LLMs experienced a substantial average increase of \$89 in weekly earnings (in 2010 prices), while unemployment exhibited only a negligible change.
These results are consistent with a complementarity story: LLMs raise workers’ productivity, thereby increasing demand for labor, and in the short run, the adjustment occurs primarily through higher wages rather than changes in employment. This pattern suggests that significant frictions in the labor market render labor supply relatively inelastic across occupations in the short term, with response to technological shocks. 





\paragraph{Related Literature}
A growing body of research examines both the potential and realized effects of AI, particularly LLMs, on the labor market. One line of work develops measures of the potential exposure of workers to AI. Early contributions include \citep{FreyOsborne2017} expert-elicited automation probabilities and \cite{BrynjolfssonMitchellRock2017}’s machine-learning suitability scores. \citep{FeltenRajSeamans2021} map AI capabilities to occupational abilities, while \citep{Webb2019} uses patent data to estimate exposure. More recent work leverages job postings and résumés to infer AI adoption. \cite{handa2025economic} propose a prompt-based measure linking O*NET tasks to millions of user prompts to Anthropic’s Claude, classifying tasks as automative, augmentative, or neither; occupational exposure is defined as the share of tasks with at least one prompt. \citep{DominskiLee2025} introduce dynamic exposure scores by querying frontier models (GPT-4o and Claude 3.5) about their ability to perform each O*NET task across five stages of technological progress. They show that exposure in occupations such as customer service rises from about 50\% under early LLMs to over 80\% once models acquire multimodal reasoning capabilities. These measures enable researchers to distinguish between automation-prone and augmentation-prone occupations and to track exposure as technology evolves. \cite{eloundou2024gpts} further assess whether LLMs meet the criteria of a general-purpose technology, quantifying affected jobs and tasks using a rubric of LLM capabilities modeled on machine-learning suitability scores, and asking both human experts and GPT-4 to evaluate the alignment between LLM abilities and hundreds of O*NET tasks.

A second line of literature tried to directly asses the contribution of LLMs to productivity productivity and work processes. Randomized controlled trials demonstrate that generative AI can substantially raise productivity in specific tasks. For instance, \cite{NoyZhang2023} and \cite{PengEtAl2023} document 15–40\% improvements in writing and coding tasks, while \citep{BrynjolfssonLiRaymond2023} find that LLM assistance boosts customer-support productivity and customer satisfaction. Finally, recent work examines directly the broader labor-market consequences of AI adoption, with mixed results.

\paragraph{LLMs, earnings and Unemployment} In our empirical analysis, we show that occupations more exposed to LLMs have experienced an increase in earnings, with little to no discernible effect on unemployment in these occupations. Similar to our work, recent studies have begun to estimate the effects of LLMs on labor market outcomes, yielding mixed results.  \cite{HumlumVestergaard2025}, using survey data from Danish workers across 11 occupations, find no significant effects on earnings or hours, although he did find an increase productivty. In contrast, \citep{BrynjolfssonChandarChen2025} exploit high-frequency payroll records from ADP and show that employment declines are concentrated among entry-level workers in highly AI-exposed occupations: employment among workers aged 22–25 in the most exposed quintile fell by around 6\% between late 2022 and mid-2025, while older workers in the same occupations experienced employment gains. They also find that declines occur where AI automates tasks but not where it augments work, and that wages change little across exposure quintiles. Similarly, \citep{DominskiLee2025} link their dynamic exposure scores to the Current Population Survey and report that increases in AI exposure are associated with lower employment, higher unemployment, and shorter work hours. On the other hand, \citep{JohnstonMakridis2025} adopt a difference-in-differences design at the sector level and find that sectors more exposed to LLMs experience significant wage and employment gains, particularly among young and highly educated workers. 

\paragraph{Road Map} In section $\ref{sec:data}$ we describe the data used for our analysis. In section \ref{sec:empircalStrat} we discuss our empirical strategy. Section \ref{sec:results} shows our empirical patterns in our data and main results 
and section \ref{sec:conclusion} concludes.

\section{Data}\label{sec:data}
For our analysis, we combine data on unemployment and earnings with measures of occupational exposure to LLMs. 

\paragraph{Earnings and Unemployment} 
To measure occupational outcomes, we use data from the Current Population Survey (CPS) from the Integrated Public Use Microdata Series (IPUMS). Occupations are classified using the Census Bureau’s 2010 occupation codes, which allows us to align them with the exposure measures from \cite{handa2025economic}, as discussed below. To obtain earnings at the occupational level, we use the CPS Outgoing Rotation Group (ORG) earnings data. The ORG includes individuals in their fourth and eighth interview months, for whom additional earnings questions are asked about their weekly earning. This enables us to construct a time series of average earnings across occupations. The reported income in the IPUMS CPS data is top-coded earnings. The top-code changed in April 2024, from a fixed threshold of 2884 to the mean of the top 3\% of earners. We adjust all earnings to real wages in January 2010 dollars. We do not apply the CPS sampling weights at the occupation level, since they do not represent within-occupation weights. Instead, we compute unweighted means within each occupation and apply weights during the analysis based on the number of observations. To measure unemployment, we use the CPS employment status variable, applying the same weighting approach.  

To measure unemployment, we classify individuals as unemployed if they are without work but actively searching for a job. We define the labor force as the sum of employed and unemployed individuals, excluding those not in the labor force. The unemployment rate for each occupation is then calculated as the number of unemployed divided by the total labor force in that occupation.  Our analysis uses monthly CPS data from January 2010 through August 2025. The mapping between O*NET tasks and CPS occupations is performed through the standard crosswalk between O*NET-SOC and Census 2010 occupation codes\footnote{\href{https://www.bls.gov/cps/cenocc2010.htm}{see here}}.

\paragraph{LLM Exposure} 
To measure occupational exposure to LLMs, we build on \cite{handa2025economic}. Specifically, they document the approximate share of Anthropic Claude prompts based on a sample of several million that are associated with each O*NET task. We define LLM exposure for an occupation as the share of its O*NET tasks that have any associated prompts. This measure reflects the fraction of potential tasks within an occupation that could be affected or altered by LLMs. In addition, \cite{handa2025economic} classify tasks as “automative,” “augmentative,” or neither. Using this classification, we construct corresponding measures of automative and augmentative exposure for each occupation. Table \ref{tab:mostCommon} reports the occupations most exposed to LLMs. These occupations rely heavily on writing and coding as their primary medium of work. Figure \ref{fig:llm_exposure} presents the distribution of exposure across occupations, with the red line indicating the median. The mode of the distribution is at zero, while the remainder is concentrated at relatively low values.

A key concern with our exposure metric is that it reflects the potential applicability of LLMs to occupational tasks rather than direct evidence of realized adoption. Because our measure is constructed from the mapping between O*NET tasks and LLM prompt distributions, it may capture a technological frontier of what LLMs could do, not necessarily what firms and workers are implementing at each period. This raises the possibility that our estimates resemble an “intent-to-treat” effect, with actual adoption rates varying across contexts. At the same time, this approach offers important advantages. First, it avoids endogeneity that would arise if we measured exposure from realized adoption, which is likely correlated with contemporaneous wage and employment shocks. Second, by anchoring the measure in task-level capabilities, we capture heterogeneity in exposure that is plausibly exogenous to short-run labor market outcomes.

\section{Empirical Strategy}\label{sec:empircalStrat}

We study the impact of LLMs on employment and earnings through a panel of occupations by month-year. Our strategy begins with the canonical difference-in-differences (DiD) design and extends it with the synthetic difference-in-differences (SDiD) estimator of \cite{arkhangelsky2021synthdid} to address potential deviations from parallel trends, driven by latent factor.

\subsection{Baseline Difference-in-Differences}

Let $i$ index occupations and $t$ time periods. Outcomes of interest $Y_{it}$ are either weekly earnings or the unemployment rate. We define treatment as
\[
W_{it} \;=\; \mathbf{1}\{t \geq T_0\}\cdot \mathbf{1}\{i \in \mathcal{T}\},
\]

where $T_0$ denotes the introduction date of ChatGPT at the end of November 2022, and $\mathcal{T}$ denotes the set of occupations classified as exposed to LLMs, defined as those with exposure above the median.

A potential concern in our design is the assumption that November 2022, coinciding with the public release of ChatGPT, marks a common treatment date across occupations. In practice, adoption of LLMs was gradual and uneven: some firms and workers experimented with these tools earlier, while others integrated them only months later. Our analysis should therefore be interpreted as capturing the average short-run response to the availability of LLMs, rather than the exact timing of adoption in every occupation. In this approach, the sudden public release of ChatGPT represents a salient, discontinuous shift in accessibility that plausibly altered expectations and behavior broadly, even among workers who did not immediately adopt. Nonetheless, our results should be understood as reflecting an “exposure shock” associated with LLM availability, not a precise measure of realized adoption intensity.

The baseline two-way fixed-effects specification is
\[
Y_{it} \;=\; \alpha_i \;+\; \delta_t \;+\; \beta\, W_{it} \;+\; \varepsilon_{it},
\]
with occupation fixed effects $\alpha_i$ and time fixed effects $\delta_t$. Let $Y_i(0)$ and $Y_i(1)$ be the potenital outcome for occupations with low exposure and high exposure respectivlty, then The coefficient $\beta$ identifies the average treatment effect on the treated (ATT), $\mathbb{E}\!\big[Y_i(1) - Y_i(0)|i \in \mathcal{T}\big]$, under the \emph{parallel trends assumption}:
\[
\mathbb{E}\!\big[ Y_{it}(0) - Y_{is}(0) \,\big|\, i \in \mathcal{T}\big] \;=\; 
\mathbb{E}\!\big[ Y_{jt}(0) - Y_{js}(0) \,\big|\, j \in \mathcal{C}\big]
\quad \forall t,s \leq T_0,
\]
where $\mathcal{C}$ denotes controls. This condition requires that, absent treatment, treated and control occupations would evolve in parallel after removing additive effects.

To assess the plausibility of this condition and to capture dynamic treatment effects, we also estimate an event-study model:
\[
Y_{it} \;=\; \alpha_i \;+\; \delta_t \;+\; 
\sum_{k \neq -1} \beta_k\,\mathbf{1}\{t - T_0 = k\}\cdot \mathbf{1}\{i \in \mathcal{T}\} \;+\; \varepsilon_{it},
\]
normalizing $\beta_{-1}=0$. Here, $\beta_k$ traces relative-time effects. Pre-treatment coefficients ($k<0$) provide descriptive evidence on whether treated and control occupations exhibited differential pre-trends. In practice, however, even small but systematic deviations can generate substantial bias in $\beta$, and conditioning inference on ``passing'' a pre-trend test worsens this bias (\cite{roth2022pretest}). Instead, we treat the event study descriptively and use evidence of drift in pre-period coefficients as motivation for an estimator that directly adjusts for such violations.

\subsection{Synthetic Difference-in-Differences}

To mitigate bias from pre-existing differential trends, we employ the SDiD estimator. SDiD constructs reweighted versions of the treated units, using the control groups units, to match the pre-trend in the pre-treatment period. Then it applies a two-way fixed effects regression to this reweighted panel. Specifically, it learns two sets of weights; \textbf{Unit weights} $\omega = (\omega_1,\dots,\omega_{N_{\mathrm{co}}})$,  which, similar to synthetic control, are assigned to control occupations, chosen to match the average pre-treatment path of treated units, and \textbf{Time weights} $\lambda = (\lambda_1,\dots,\lambda_{T_{\mathrm{pre}}})$, assigned to pre-treatment periods, chosen to emphasize the components of the pre-period most predictive of the post-period for controls.

Formally, let $\bar{Y}_{\mathcal{T},t}$ denote the average outcome of the treated group at time $t$, and $\bar{Y}_{\mathcal{C},t}(\omega) = \sum_{j\in \mathcal{C}} \omega_j Y_{jt}$ the weighted average of controls. The \emph{unit weights} are chosen by solving
\[
\widehat{\omega} \;=\; \arg\min_{\omega \geq 0, \; \sum_j \omega_j = 1} 
\sum_{t \leq T_0}\, \big(\bar{Y}_{\mathcal{T},t} - \bar{Y}_{\mathcal{C},t}(\omega)\big)^2,
\]
so that the synthetic control of controls replicates the pre-period trajectory of treated units.

The \emph{time weights} are chosen by solving an analogous problem across time,
\[
\widehat{\lambda} \;=\; \arg\min_{\lambda \geq 0, \; \sum_{t\leq T_0} \lambda_t = 1} 
\sum_{j \in \mathcal{C}} \bigg(\sum_{t \leq T_0} \lambda_t Y_{jt} - \frac{1}{T_{\mathrm{post}}} \sum_{t > T_0} Y_{jt}\bigg)^2,
\]
so that the relevant linear combination of pre-periods best predicts the control group’s post-period outcomes.

Given these weights, \cite{arkhangelsky2021synthdid} shows that the ATT can be estimated as the coefficient $\tau$ in the weighted regression:
\[
(\widehat{\tau}, \widehat{\mu}, \widehat{\alpha}, \widehat{\beta}) \;=\;
\arg\min_{\tau,\mu,\alpha,\beta}\; 
\sum_{i,t}\, \widehat{\omega}_i\,\widehat{\lambda}_t\,
\big( Y_{it} - \mu - \alpha_i - \beta_t - \tau W_{it} \big)^2.
\]

The SDiD framework can be interpreted within an interactive fixed-effects model for untreated potential outcomes:
\[
Y_{it}(0) \;=\; \alpha_i \;+\; \delta_t \;+\; \gamma_i^{\top} \upsilon_t \;+\; \varepsilon_{it}.
\]
Here, $\gamma_i^{\top}\upsilon_t$ represents latent factors driving persistent heterogeneity in trends across occupations. Standard DiD requires that differencing removes this component; equivalently, it assumes $\gamma_i$ does not vary systematically between treated and controls. SDiD relaxes this assumption by using pre-treatment information to balance the factor structure. Unit weights $\omega$ align the span of $\gamma_i$ between treated and synthetic controls, while time weights $\lambda$ ensure that the relevant linear combinations of $\upsilon_t$ in the pre-period approximate the post-period trajectory. Under mild regularity conditions, this ``double balancing'' eliminates bias from the latent factor component.

In this sense, SDiD is robust to "factor style" violations of parallel trends. It does not require that differencing alone suffices, nor that a single sparse synthetic control exactly replicates the treated path. Instead, it achieves balance across both units and time, restoring the comparability of counterfactuals at the margin relevant for post-treatment inference.

Finally, the SDiD can be estimated in two ways: either simultaneously across all treated units or separately for each treated unit. In our empirical analysis, we adopt the latter approach. Specifically, we estimate the treatment effect for each treated unit individually and then report both the average treatment effect across units and the distribution of these unit-level effects.

There are two main reasons for this choice. First, the SDiD estimator requires a balanced panel in order to be implemented. This requirement implies that occupations with relatively few observations may be excluded when the estimation is carried out jointly across all treated units. By estimating treatment effects unit by unit, we can overcome this limitation. For each treated unit, we restrict the donor pool to control units that form a balanced panel with it, thereby retaining more occupations in the analysis. Second, estimating effects unit by unit allows us to examine heterogeneity in treatment effects across occupations. This perspective is particularly important in our setting, since the exposure measure is binarized and the effects are likely to be heterogeneous.

\section{Results}\label{sec:results}
\subsection{Descriptive Patterns}
\begin{figure}[H]
    \centering
    \begin{minipage}[b]{0.45\textwidth}
        \centering
        \textbf{Unemployment}\\[0.5em]\includegraphics[width=\textwidth]{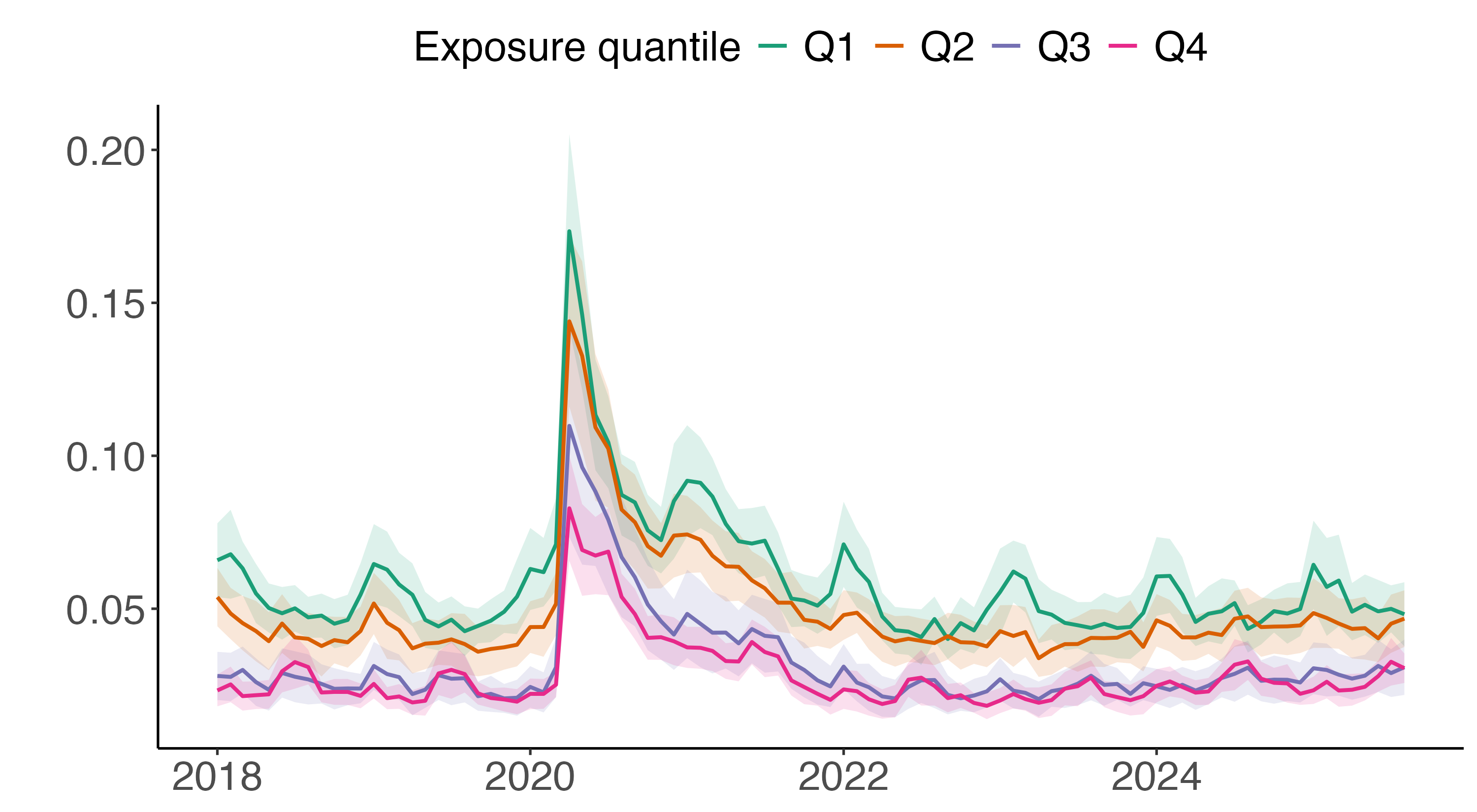}
        \caption{Unemployment by Quartile of Exposure}
        \label{fig:trends_unemploymemnt}
    \end{minipage}
    \hfill
    \begin{minipage}[b]{0.45\textwidth}
        \centering
        \textbf{Earnings}\\[0.5em]\includegraphics[width=\textwidth]{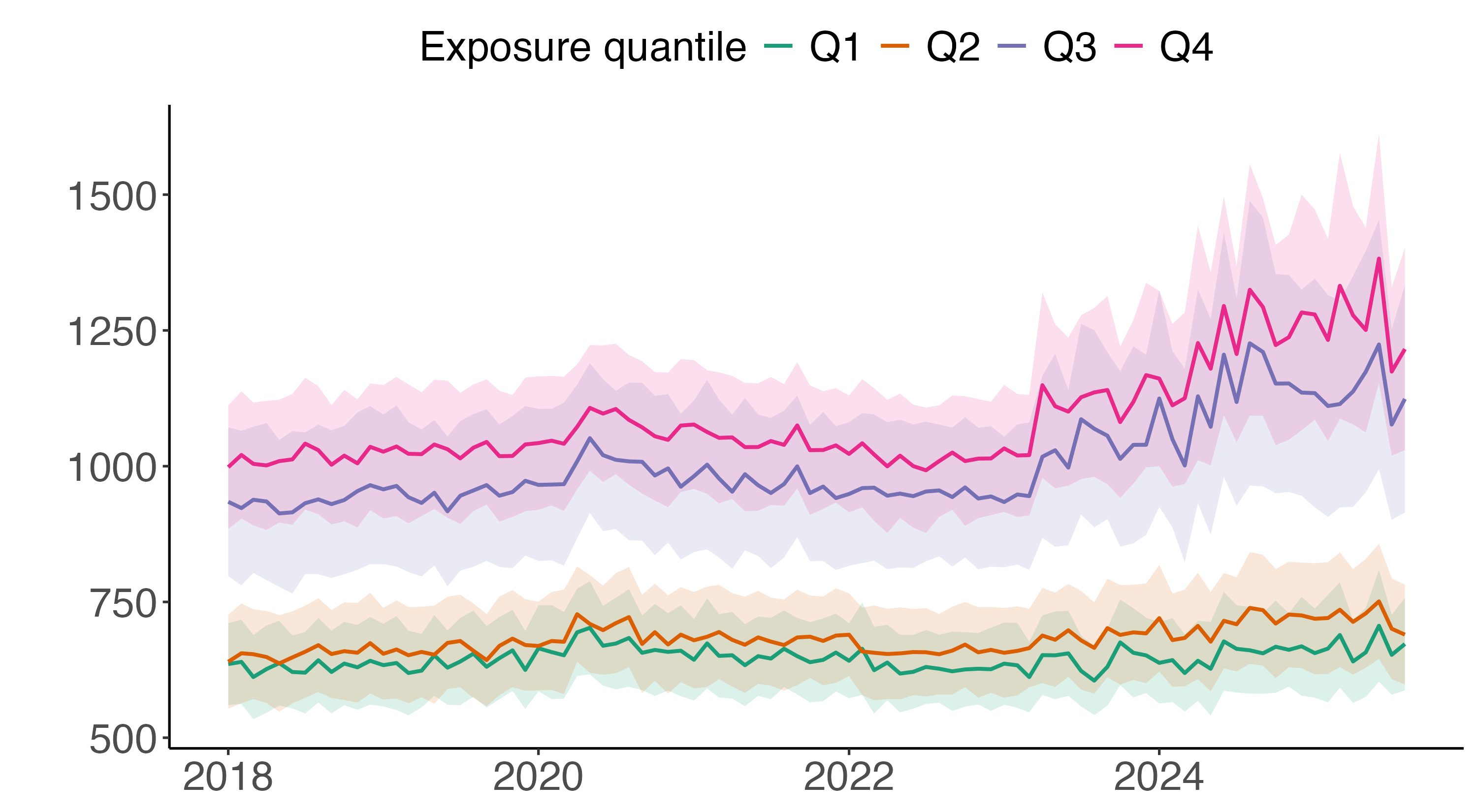}
        \caption{Average Weekly Earnings by Quartile of Exposure}
        \label{fig:trends_earnings}
    \end{minipage}\\
    \justifying
    \footnotesize\textit{ 
    Note: Figure 1 shows unemployment rates by quartile of LLM exposure, calculated from CPS monthly data between January 2010 and August 2025. Each line corresponds to one quartile of occupations ranked by their exposure measure. Figure 2 shows average weekly earnings over the same period, again split by quartiles of exposure, with earnings adjusted to January 2010 dollars.  The shaded area indicates the 95\% confidence interval, computed using clustered standard errors.}
\end{figure}

We begin by examining the raw trends in unemployment and earnings across occupations, grouped into four quartiles by their level of exposure to LLMs. Figure~\ref{fig:trends_unemploymemnt} displays unemployment trends from January 2010 through August 2025. Overall, unemployment rates appear relatively stable across quartiles, with the exception of the temporary spike during the Covid-19 pandemic. Notably, occupations that are ex-post more exposed to LLMs consistently exhibit lower unemployment throughout the sample period: in the two years prior to the introduction of ChatGPT, average unemployment was 4.1\% for treated occupations compared to 5.8\% for controls.

Figure \ref{fig:trends_earnings} plots average weekly earnings over the same horizon. Occupations in higher exposure quartiles generally earn more than those in lower quartiles, but the slopes of the pre-treatment earnings trajectories are visually similar across groups. In the post-treatment period, and particularly in 2024 and 2025, we observe a sharp increase in earnings for occupations in the third and fourth quartiles of exposure. Between December 2022 and August 2025, occupations in the top two quartiles experienced an average increase of \$105 in weekly earnings, compare to the lower quintiles experiencing a rise of approximately \$60. 

Figures \ref{fig:trends_unemploymemnt_aug} and \ref{fig:trends_earnings_aug} in the Appendix replicate the analysis using alternative exposure measures based on the share of tasks subject to augmentation. Similarly, Figures \ref{fig:trends_earnings_automation} and \ref{fig:trends_unemploymemnt_automation} present the corresponding exercise for exposure to automation. Both sets of results display a pattern similar to the one shown here.

\subsection{Difference-in-Differences Analysis}

\begin{figure}[htbp]
    \centering
    \begin{minipage}[b]{0.45\textwidth}
        \centering
        \textbf{Unemployment}\\[0.5em] \includegraphics[width=\textwidth]{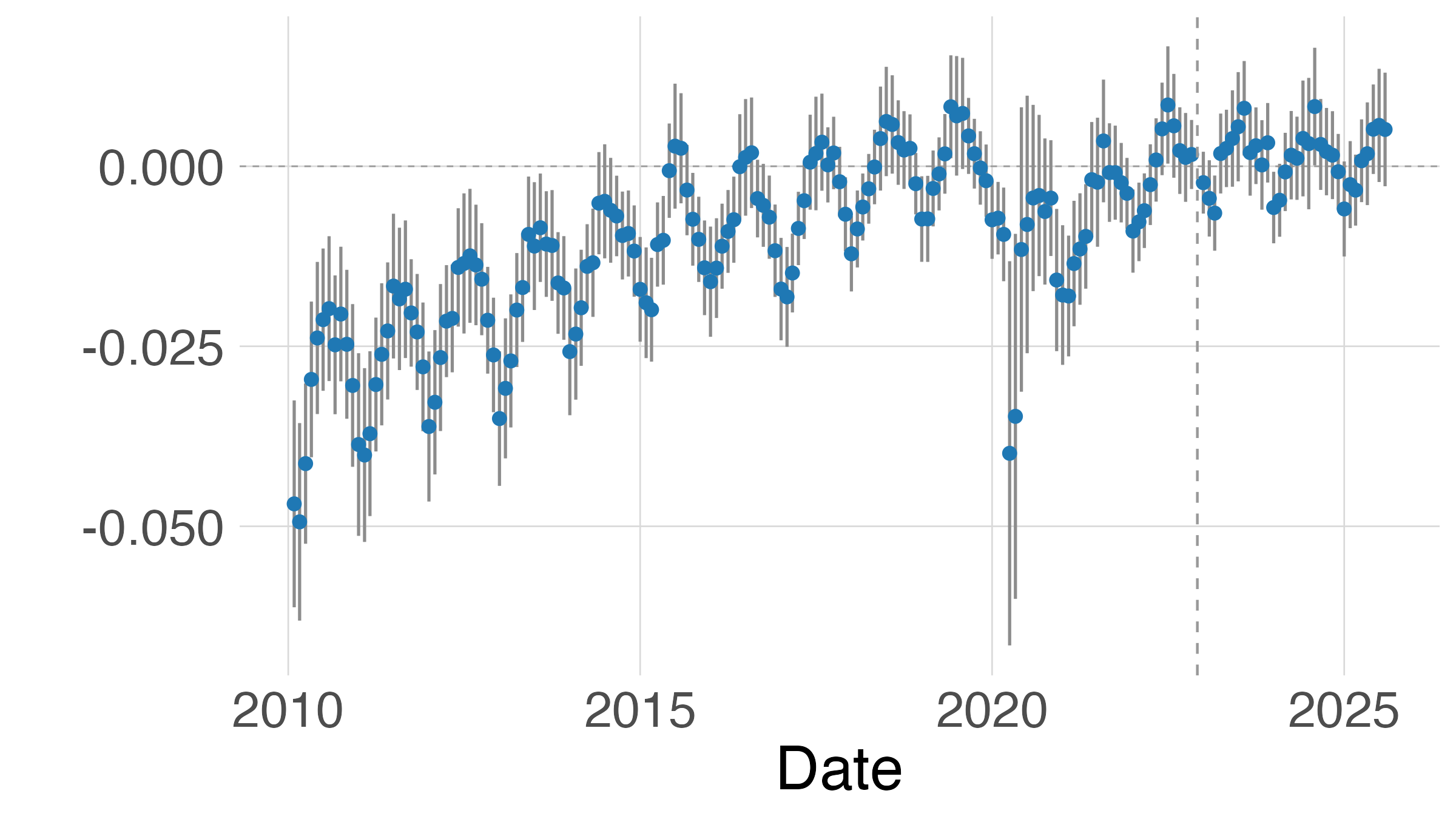}
        \caption{Event Study --- Unemployment}
        \label{fig:did_pretrends_unemployment}
    \end{minipage}
    \hfill
    \begin{minipage}[b]{0.45\textwidth}
        \centering
        \textbf{Earnings}\\[0.5em]\includegraphics[width=\textwidth]{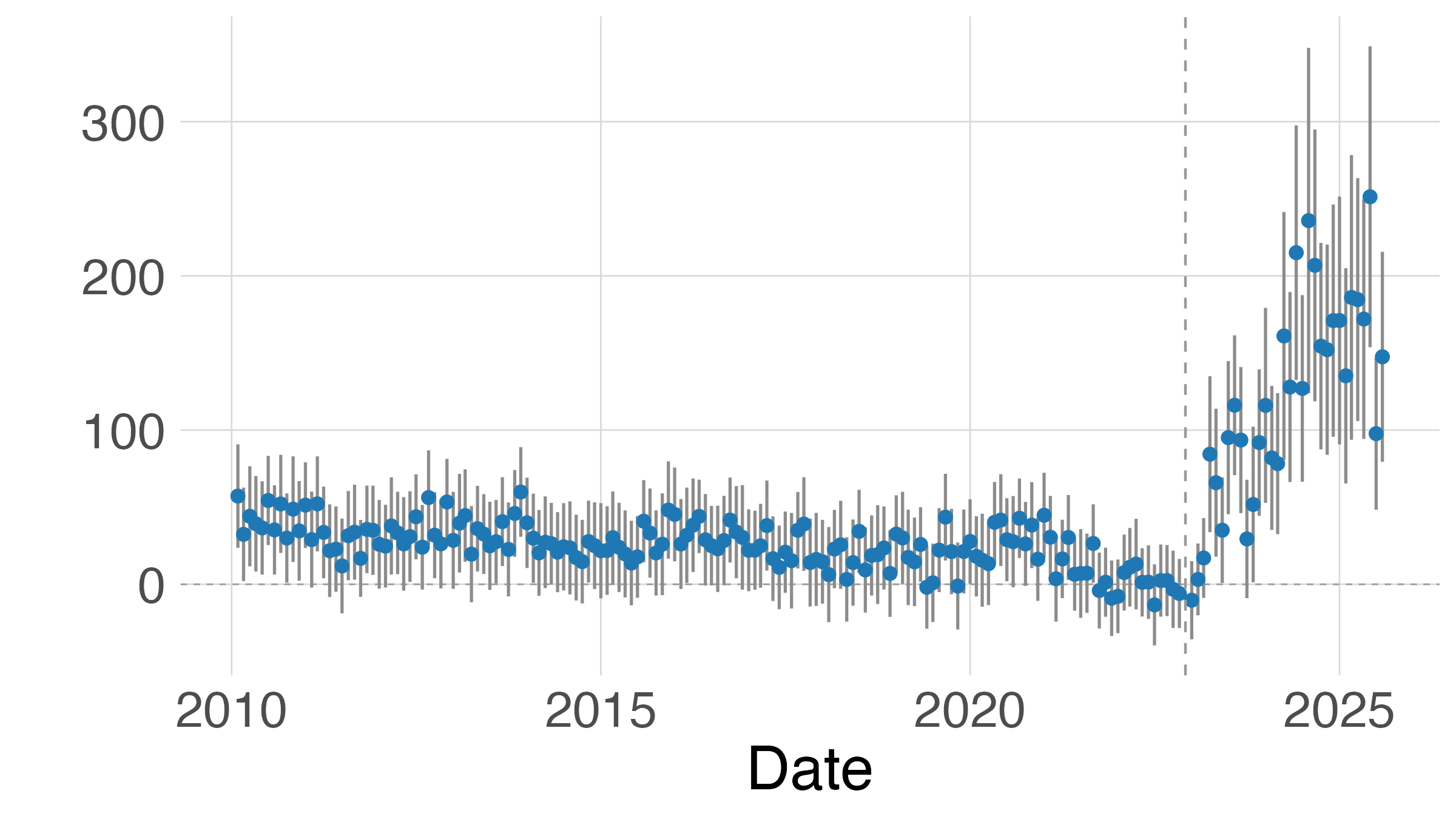}
        \caption{Event Study --- Earnings}
        \label{fig:did_pretrends_earnings}
    \end{minipage}
    \\
    \justifying
    \footnotesize\textit{ 
    Note: Figure 3 presents event-study estimates of unemployment rates for occupations with high-exposure (above the median exposure rate) versus low-exposure, relative to the introduction of ChatGPT in November 2022 (normalized to 0 at one period before treatment). Figure 4 presents the same event-study specification for average weekly earnings. Coefficients are shown relative to the pre-treatment baseline, with separate lines for treated and control occupations. The gray lines indicates the 95\% confidence interval, computed using clustered standard errors.}
\end{figure}

Figures \ref{fig:did_pretrends_unemployment} and \ref{fig:did_pretrends_earnings} present event-study plots for unemployment and earnings, respectively. We define treated occupations as occupations with above median exposure to LLM. For unemployment, figure \ref{fig:did_pretrends_unemployment} shows a clear violations of the parallel-trend assumption, at the pre-period, required for DiD to yield causal estimates. Treated and control occupations display differential dynamics even before the introduction of LLMs, making the parallel-trends assumption implausible. By contrast, the pre-trend for earnings appears more stable: although some deviations emerge in earlier years, the differences are smaller and less systematic.

In line with the raw trends in Figure \ref{fig:trends_earnings}, the post-treatment period shows a widening earnings gap between high-exposure and low-exposure occupations. Estimates from the baseline DiD specifications are presented in Table \ref{tab:treatment_estimates} in the Appendix. The results show sizable positive effects of LLM exposure on weekly earnings and unemployment of around \$95 for weekly earnings, and 1.2 percentage points for unemployment; however, given the observed pre-trends, these effects are unlikely to be causal.

Figures \ref{fig:did_pretrends_unemployment_auto} -\ref{fig:did_pretrends_earnings_aug} in the Appendix replicate the analysis using alternative exposure measures based on the share of tasks subject to augmentation versus automation, yielding similar patterns. Appendix Figures \ref{fig:did_pretrends_earnings_cont} and \ref{fig:did_pretrends_unemployment_cont} further demonstrate robustness to using the continuous exposure measure rather than a binary treatment indicator.

\subsection{Synthetic Difference-in-Differences Results}

\begin{figure}[htbp]
    \centering
    \begin{minipage}[b]{0.45\textwidth}
        \centering
        \textbf{Unemployment}\\[0.5em]
        \includegraphics[width=\textwidth]{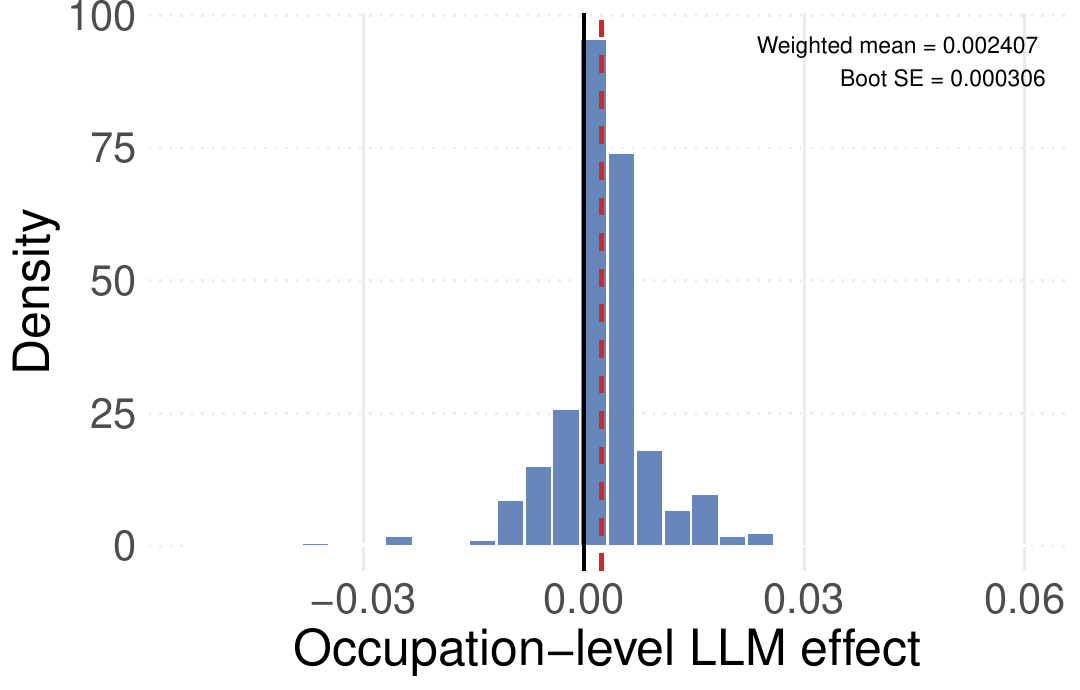}
        \caption{Distribution of SDiD Estimates: Unemployment}
        \label{fig:SDiD_unemployment}
    \end{minipage}
    \hfill
    \begin{minipage}[b]{0.45\textwidth}
        \centering
        \textbf{Earnings}\\[0.5em]\includegraphics[width=\textwidth]{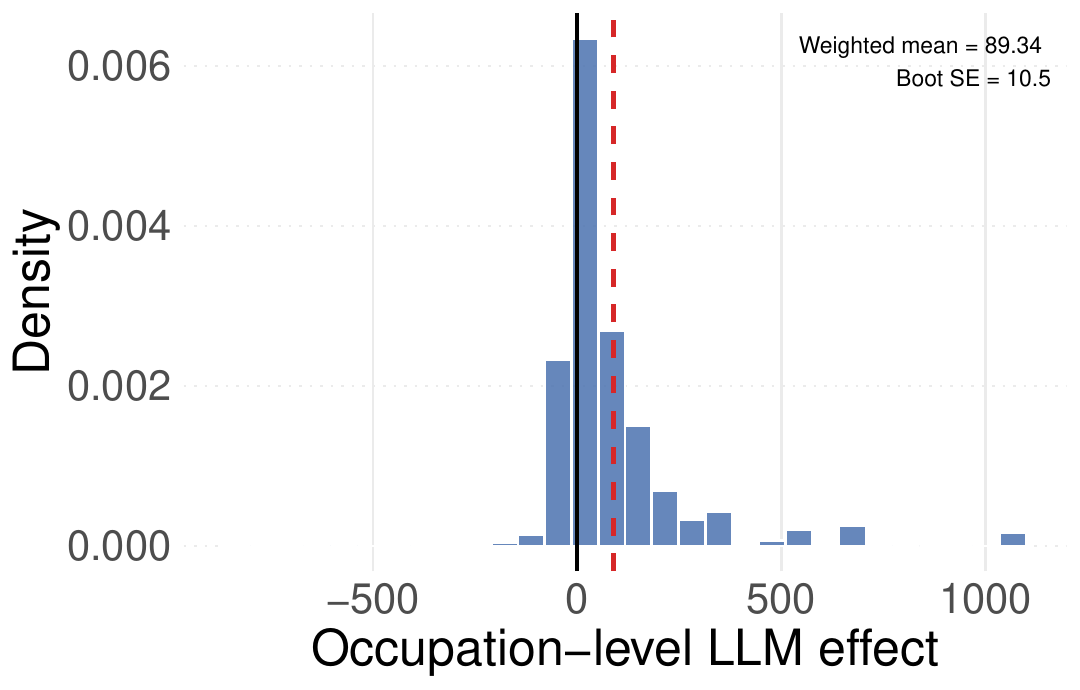}
        \caption{Distribution of SDiD Estimates: Weekly Earnings}
        \label{fig:SDiD_earnings}
    \end{minipage}\\
    \justifying
    \footnotesize\textit{Note: Figure 5 shows a histogram of Synthetic Difference-in-Differences (SDiD) estimates of the effect of LLM exposure on unemployment at the occupation level. Each bar represents the number of treated occupations with a given estimated effect. The horizontal axis reports the estimated change in the unemployment rate (percentage points), and the vertical axis shows the density across occupations. The red vertical line indicates the weighted average treatment effect on the treated (ATT). Figure 6 shows the corresponding histogram of SDiD estimates for weekly earnings at the occupation level. The horizontal axis reports the estimated change in real weekly earnings (January 2010 dollars), and the vertical axis shows the density of occupations. The red vertical line indicates the weighted average ATT, representing the mean earnings effect across treated occupations. The standard errors of the ATT are computed using a bootstrap procedure with 1,000 repetitions.}
\end{figure}

Figure~\ref{fig:SDiD_unemployment} presents the distribution of estimated causal effects from the SDiD model across treated occupations. For unemployment, the estimated effects are tightly centered around zero. While some occupations experienced modest increases and others decreases, the overall mean effect is just 0.2 percentage points. This suggests that, on average, LLM exposure did not induce systematic changes in unemployment rates at the occupation level, even though certain occupations displayed larger idiosyncratic responses.

Figure \ref{fig:SDiD_earnings} shows the corresponding distribution of effects on weekly earnings across treated occupations. In contrast to unemployment, the bulk of the distribution lies on the positive side, with an average effect of approximately \$89 per week. Although a subset of occupations experienced declines in earnings, the predominance of treated occupations have experienced a positive effects. The dispersion in effects highlights that LLMs are unlikely to exert uniform impacts across occupations and jobs: for some they raise earnings, while for others they may reduce them. This heterogeneity is consistent with the idea that LLM adoption interacts differently with occupation-specific tasks and skill demands.

Appendix Figure \ref{fig:SDiD_unemployment_aug}-\ref{fig:SDiD_earnings_aut} extends the analysis by separating exposure into augmentation-related and automation-related tasks, yielding similar patterns of heterogeneous, but predominantly positive earnings effects.
  
\section{Conclusion}\label{sec:conclusion}

This paper provides early evidence on the effects of LLMs on the labor market. Our findings suggest that the primary margin of adjustment operates through wages rather than employment. Occupations more exposed to LLMs experience notable earnings gains, consistent with productivity improvements that increase labor demand. By contrast, we find no systematic changes in unemployment, which remains relatively low for exposed occupations both before and after LLM adoption. Taken together, these results point to short-run productivity gains that translate into higher wages, while the quantity of labor. This pattern is consistent with a setting in which labor supply is not sufficiently elastic to accommodate an abrupt productivity-induced demand shifts.

\newpage
\appendix
\renewcommand{\thefigure}{A\arabic{figure}}
\setcounter{figure}{0} 

\section{Appendix}

\begin{figure}[H]
  \centering
  \begin{minipage}[]{0.55\textwidth}
    \centering
    \includegraphics[width=\linewidth]{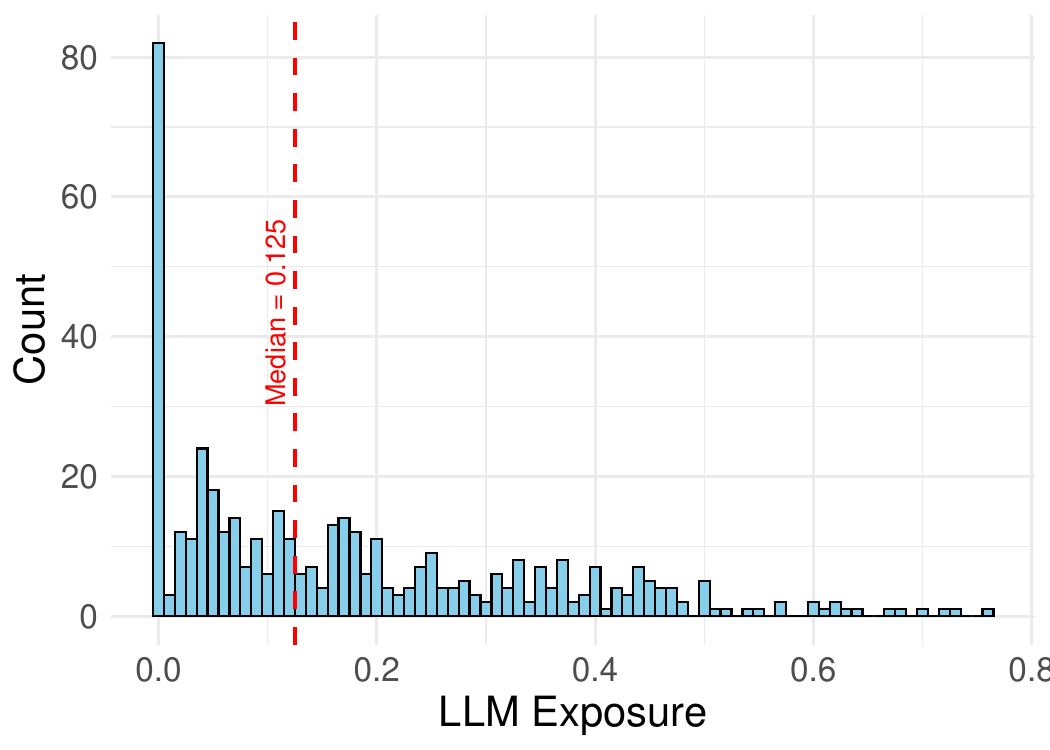}
    \caption{Distribution of LLM exposure.}
    \label{fig:llm_exposure}
  \end{minipage}\hfill
  \begin{minipage}[]{0.45\textwidth}
    \footnotesize
    \begin{tabularx}{\linewidth}{@{}>{\raggedright\arraybackslash}X@{}}
      \toprule
      Miscellaneous media and communication workers \\
      Computer programmers \\
      Writers and authors \\
      Web developers \\
      Software developers, applications and systems software \\
      Information security analysts \\
      \bottomrule
    \end{tabularx}
        \captionsetup{type=table}
        \caption{Most exposed occupations.}
    \label{tab:mostCommon}

  \end{minipage}
    \justifying
    \footnotesize\textit{ 
    Note: Figure \ref{fig:llm_exposure} presents the distribution of LLM exposure across occupations. The dashed red line marks the median of the distribution, which we use to define our treatment. Table \ref{tab:mostCommon} lists the six occupations most exposed to LLMs according to our classification.} 
\end{figure}

\begin{figure}[H]
    \centering
    \begin{minipage}[b]{0.45\textwidth}
        \centering
        \includegraphics[width=\textwidth]{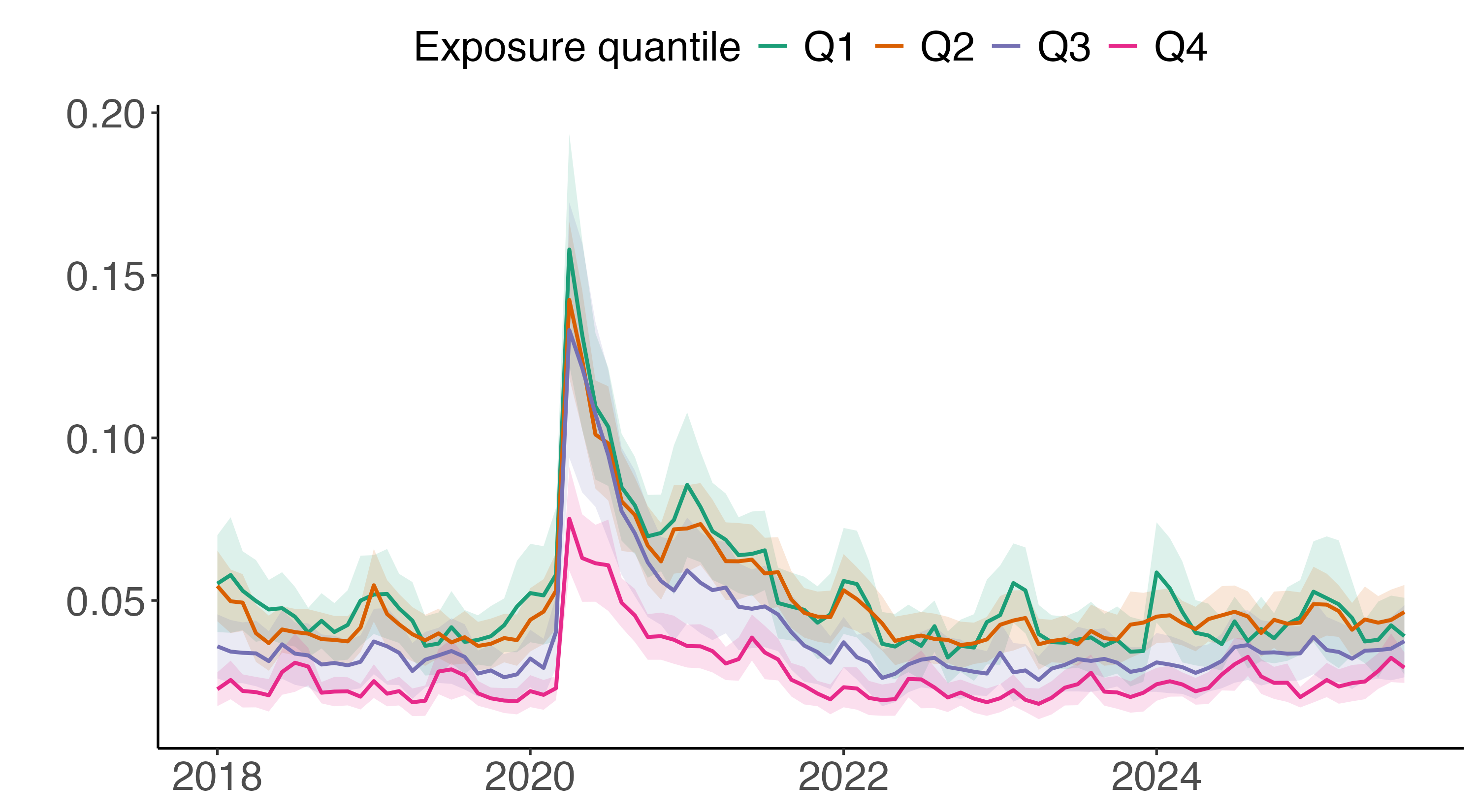}
        \caption{Unemployment by Quartile of \textbf{Augmentation} Exposure}
        \label{fig:trends_unemploymemnt_aug}
    \end{minipage}
    \hfill
    \begin{minipage}[b]{0.45\textwidth}
        \centering
        \includegraphics[width=\textwidth]{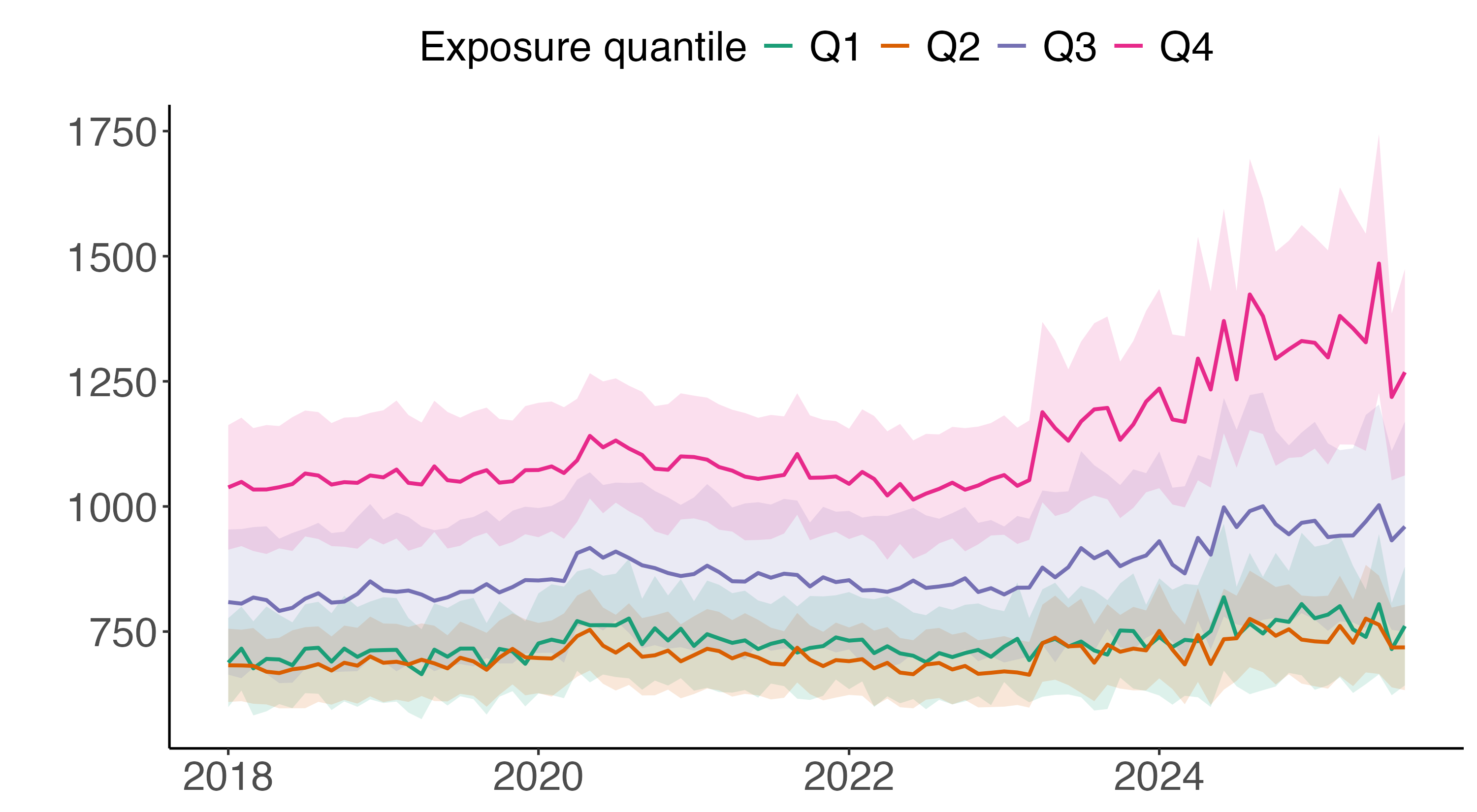}
        \caption{Average Weekly Earnings by Quartile of \textbf{Augmentation} Exposure}
        \label{fig:trends_earnings_aug}
    \end{minipage}
    \\
    \justifying
    \footnotesize\textit{ 
    Note: Figure 1 shows unemployment rates by quartile of occupations exposure to augmentation from LLM, calculated from CPS monthly data between January 2010 and August 2025. Each line corresponds to one quartile of occupations ranked by their exposure measure. Figure 2 shows average weekly earnings over the same period, again split by quartiles of exposure, with earnings adjusted to January 2010 dollars.}
\end{figure}

\begin{figure}[H]
    \centering
    \begin{minipage}[b]{0.45\textwidth}
        \centering
        \includegraphics[width=\textwidth]{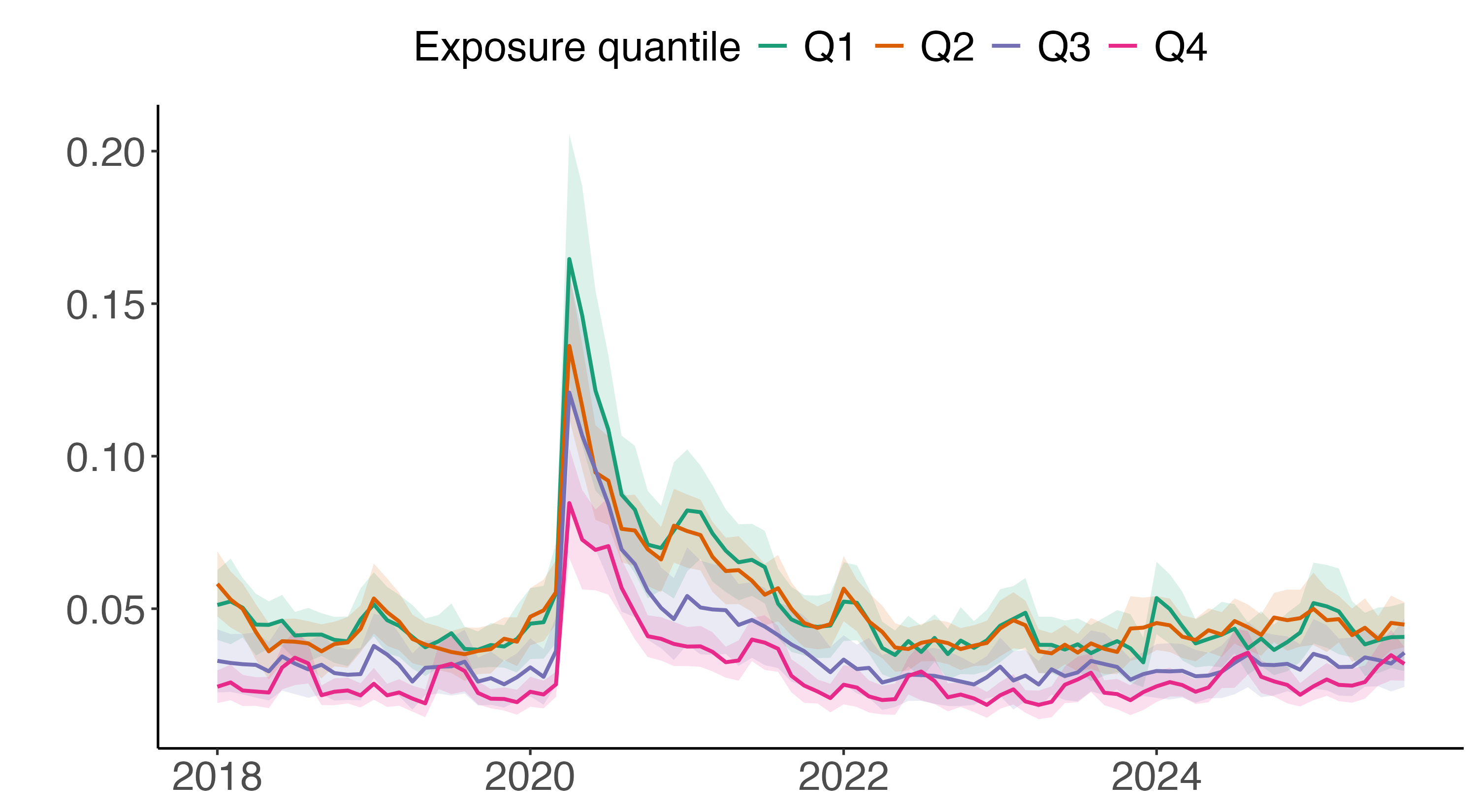}
        \caption{Unemployment by Quartile of \textbf{Automation} Exposure}
        \label{fig:trends_unemploymemnt_automation}
    \end{minipage}
    \hfill
    \begin{minipage}[b]{0.45\textwidth}
        \centering
        \includegraphics[width=\textwidth]{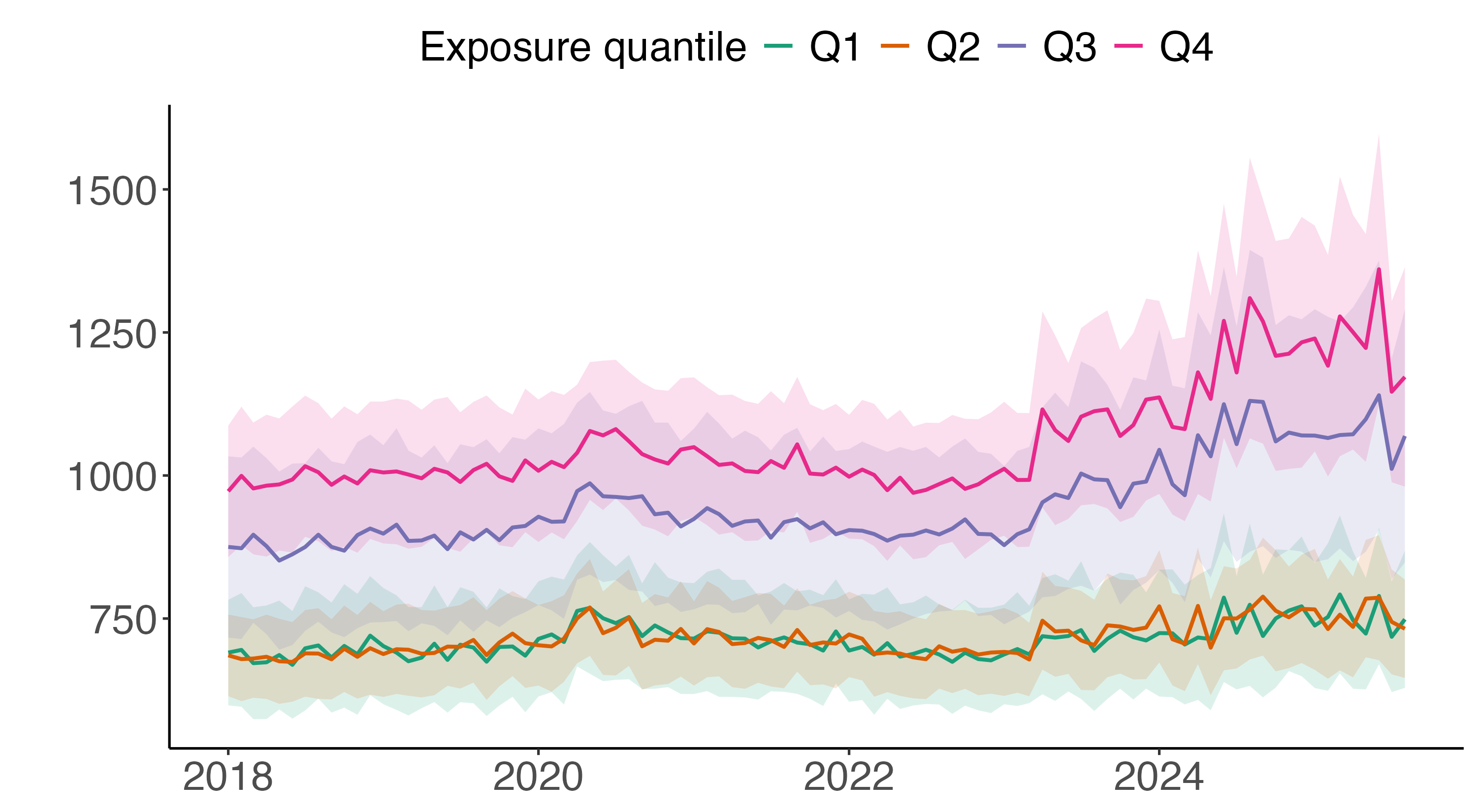}
        \caption{Average Weekly Earnings by Quartile of \textbf{Automation} Exposure}
        \label{fig:trends_earnings_automation}
    \end{minipage}
        \\
    \justifying
    \footnotesize\textit{ 
    Note: Figure 1 shows unemployment rates by quartile of occupations exposure to automation from LLM, calculated from CPS monthly data between January 2010 and August 2025. Each line corresponds to one quartile of occupations ranked by their exposure measure. Figure 2 shows average weekly earnings over the same period, again split by quartiles of exposure, with earnings adjusted to January 2010 dollars.  The shaded area indicates the 95\% confidence interval.}
\end{figure}

\begin{figure}[H]
    \centering
    \begin{minipage}[b]{0.45\textwidth}
        \centering
        \includegraphics[width=\textwidth]{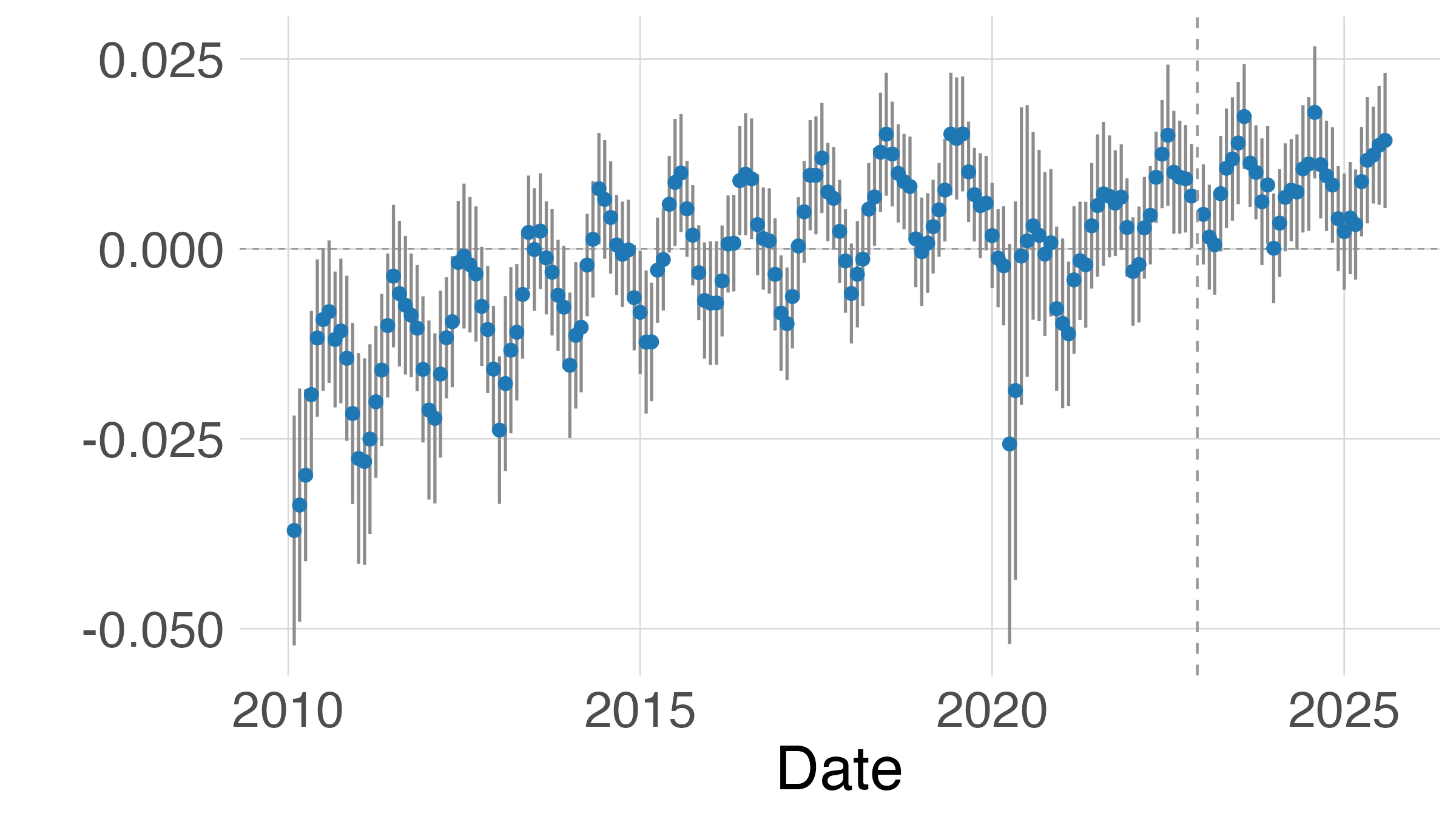}
        \caption{Event Study --- Unemployment, Exposure to Augmentation}
        \label{fig:did_pretrends_unemployment_auto}
    \end{minipage}
    \hfill
    \begin{minipage}[b]{0.45\textwidth}
        \centering
        \includegraphics[width=\textwidth]{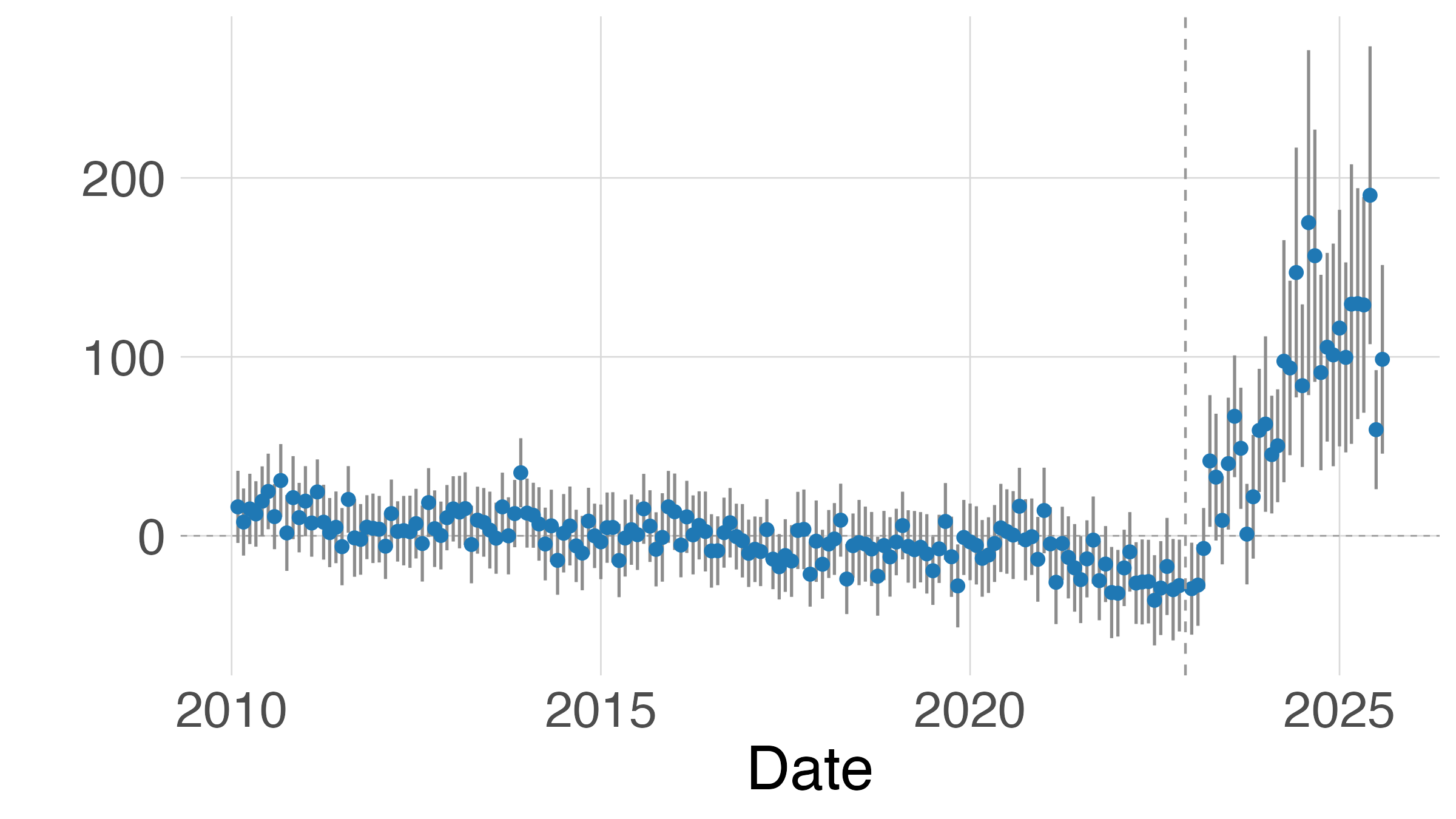}
        \caption{Event Study --- Earnings, Exposure to Automation}
        \label{fig:did_pretrends_earnings_auto}
    \end{minipage}
    \\
    \justifying
    \footnotesize\textit{ 
    Note: Figure \ref{fig:did_pretrends_earnings_auto}presents event-study estimates of unemployment rates for occupations with high-exposure (above the median exposure measured for automation tasks) versus low-exposure, relative to the introduction of ChatGPT in November 2022 (normalized to 0 at one period before treatment). Figure \ref{fig:did_pretrends_earnings_auto} presents the same event-study specification for average weekly earnings. Coefficients are shown relative to the pre-treatment baseline, with separate lines for treated and control occupations.  The gray lines indicates the 95\% confidence interval, computed using clustered standard errors.}
\end{figure}

\begin{figure}[H]
    \centering
    \begin{minipage}[b]{0.45\textwidth}
        \centering
        \includegraphics[width=\textwidth]{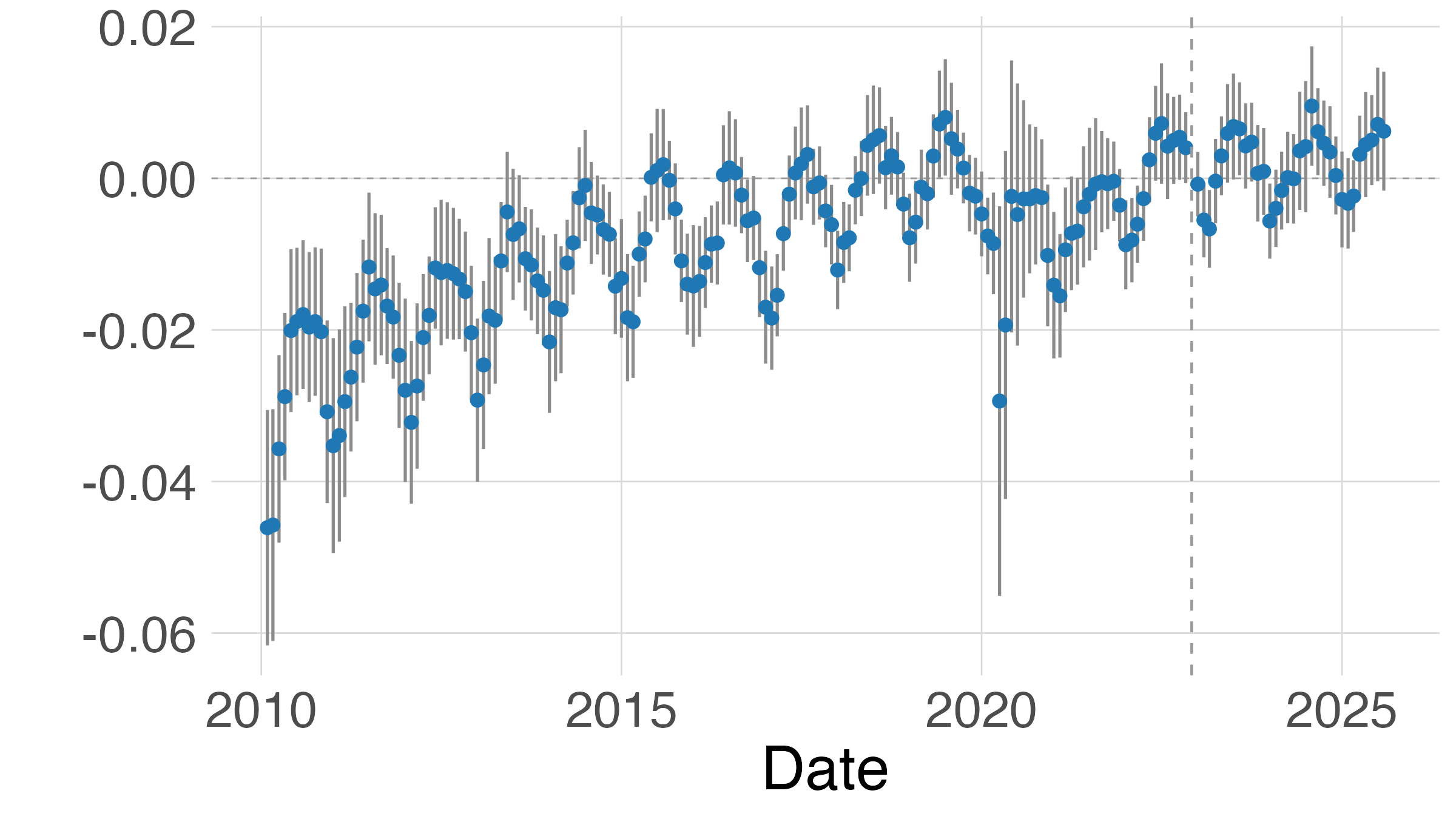}
        \caption{Event Study --- Unemployment, Exposure to Augmentation}
        \label{fig:did_pretrends_unemployment_aug}
    \end{minipage}
    \hfill
    \begin{minipage}[b]{0.45\textwidth}
        \centering
        \includegraphics[width=\textwidth]{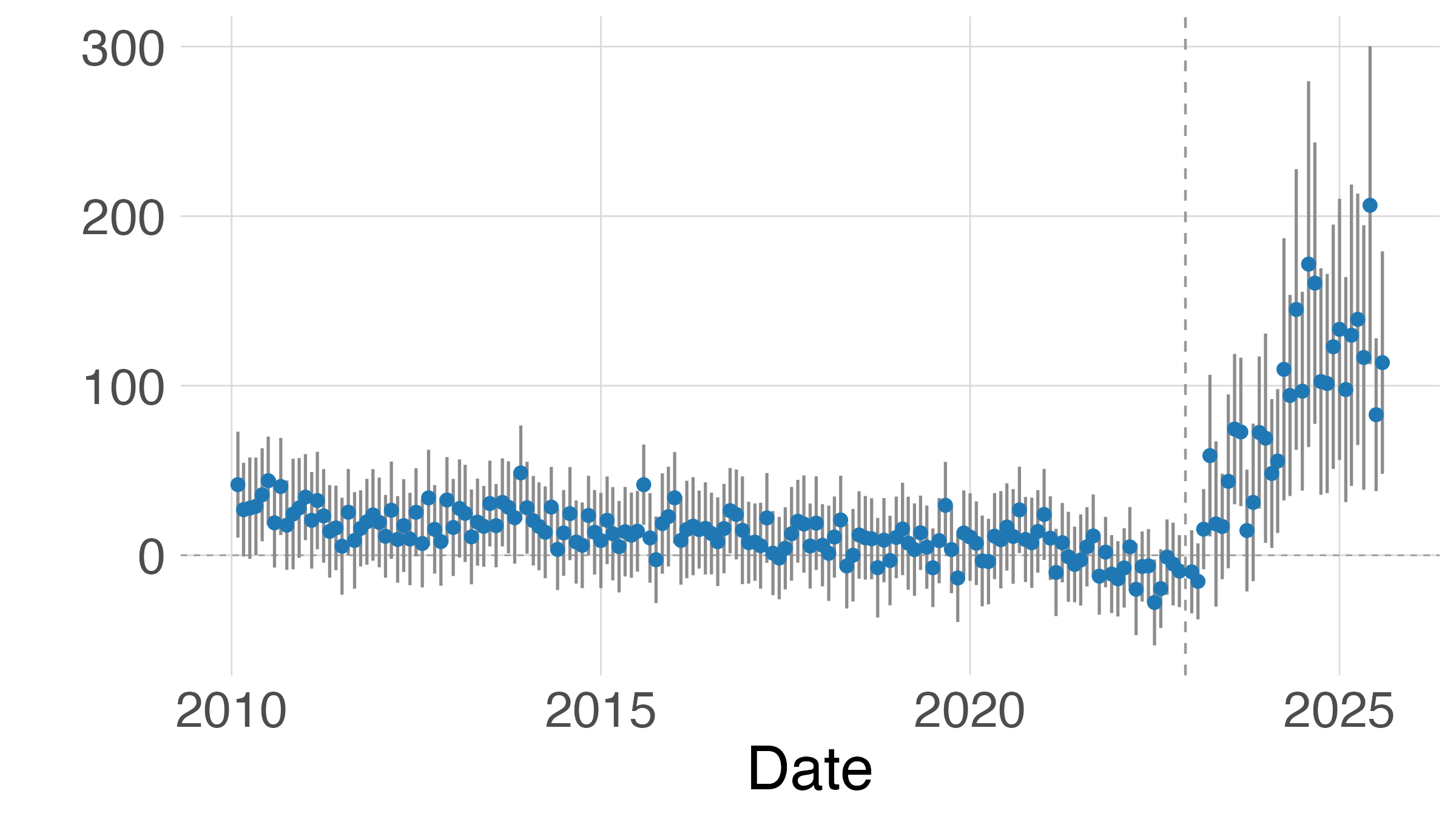}
        \caption{Event Study --- Earnings, Exposure to Automation}
        \label{fig:did_pretrends_earnings_aug}
    \end{minipage}
    \\
    \justifying
    \footnotesize\textit{ 
    Note: Figure \ref{fig:did_pretrends_earnings_aug}presents event-study estimates of unemployment rates for occupations with high-exposure (above the median exposure measured for automation tasks) versus low-exposure, relative to the introduction of ChatGPT in November 2022 (normalized to 0 at one period before treatment). Figure \ref{fig:did_pretrends_earnings_aug} presents the same event-study specification for average weekly earnings. Coefficients are shown relative to the pre-treatment baseline, with separate lines for treated and control occupations.  The gray lines indicates the 95\% confidence interval, computed using clustered standard errors.}
\end{figure}

\begin{figure}[H]
    \centering
    \begin{minipage}[b]{0.45\textwidth}
        \centering
        \includegraphics[width=\textwidth]{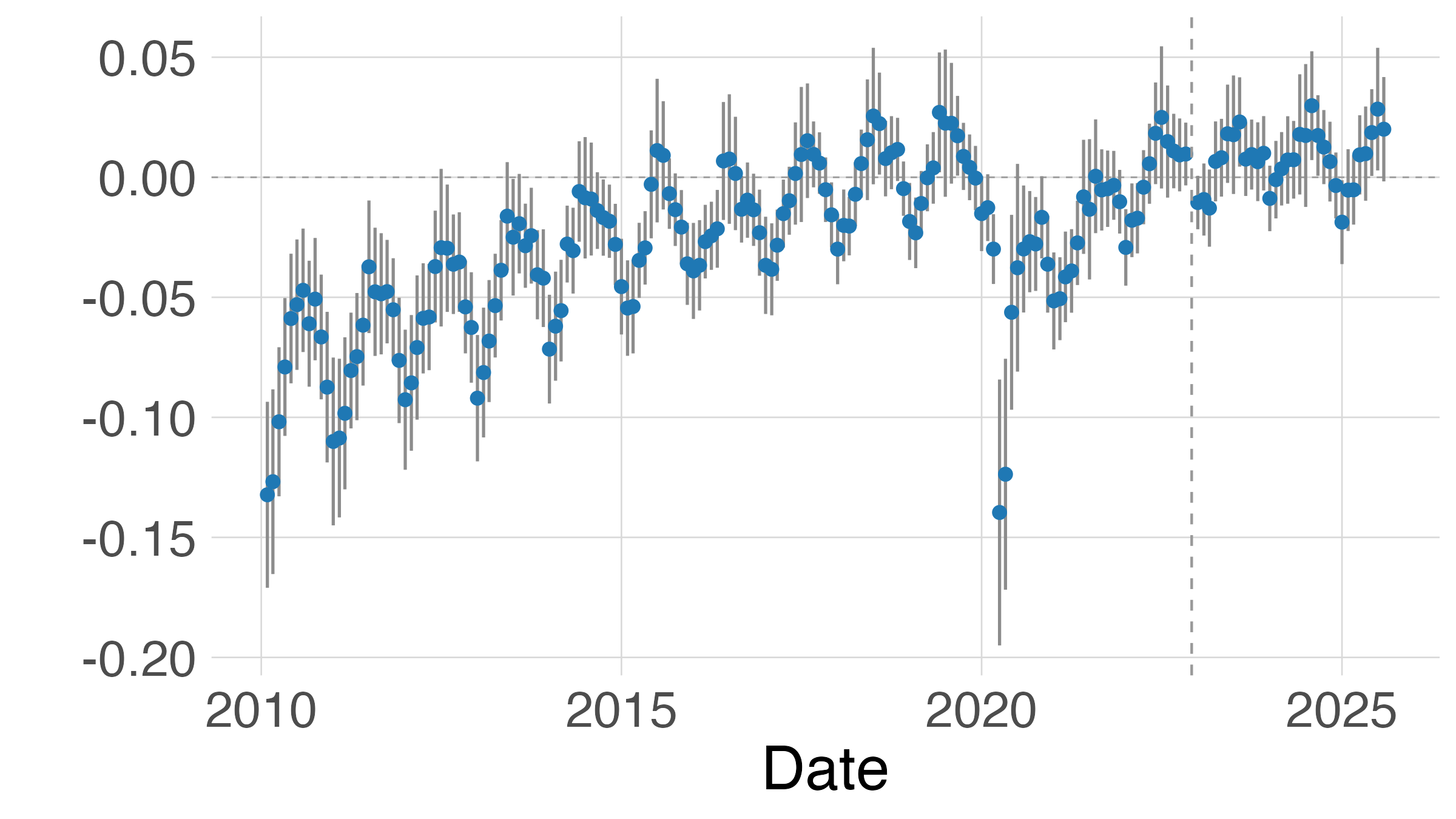}
        \caption{Event Study --- Unemployment, Continuous Treatment}
        \label{fig:did_pretrends_unemployment_cont}
    \end{minipage}
    \hfill
    \begin{minipage}[b]{0.45\textwidth}
        \centering
        \includegraphics[width=\textwidth]{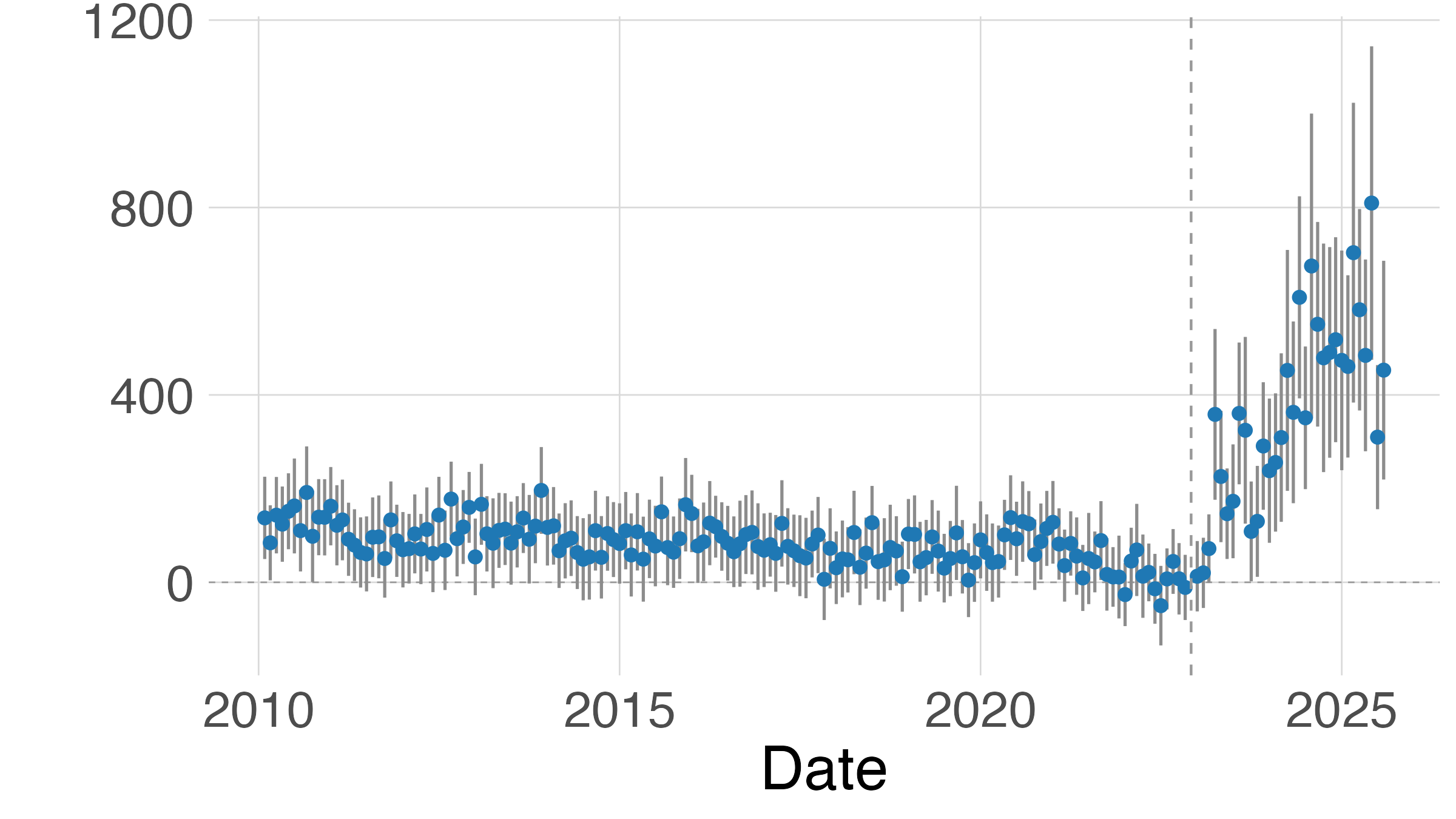}
        \caption{Event Study --- Earnings, Continuous Treatment }
        \label{fig:did_pretrends_earnings_cont}
    \end{minipage}
    \\
    \justifying
    \footnotesize\textit{ 
    Note: Figure \ref{fig:did_pretrends_unemployment_cont}presents event-study estimates of on the interaction between date variable and a continuous variable for LLM exposure as defined in section \ref{data}. Figure \ref{fig:did_pretrends_earnings_cont} presents the same event-study specification for average weekly earnings. Coefficients are shown relative to the pre-treatment baseline, with separate lines for treated and control occupations.  The gray lines indicates the 95\% confidence interval, computed using clustered standard errors.}
\end{figure}

\begin{figure}[H]
    \centering
    \begin{minipage}[b]{0.45\textwidth}
        \centering
        \includegraphics[width=\textwidth]{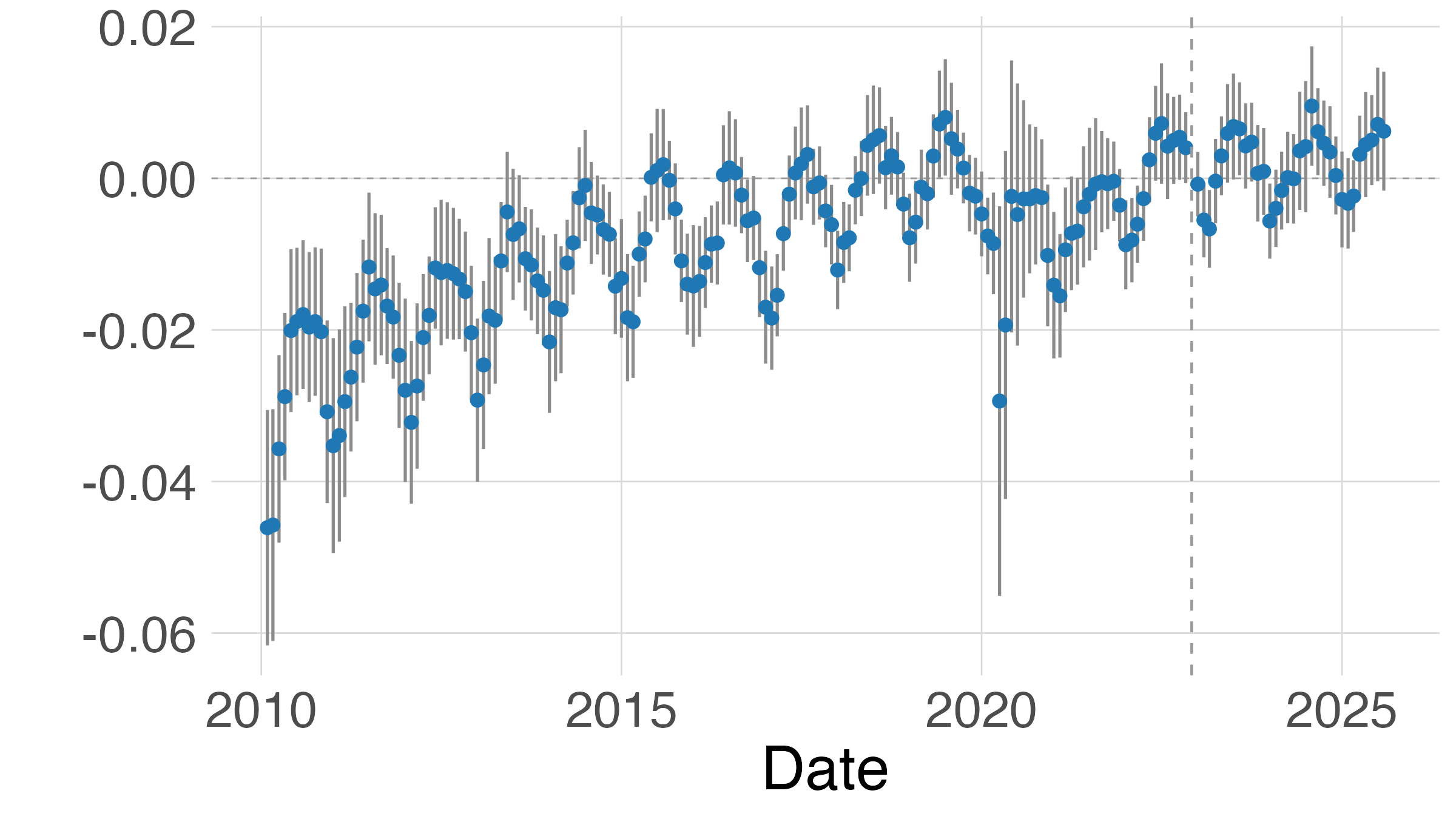}
        \caption{Event Study --- Unemployment, Exposure to Augmentation}
        \label{fig:did_pretrends_unemployment_aug}
    \end{minipage}
    \hfill
    \begin{minipage}[b]{0.45\textwidth}
        \centering
        \includegraphics[width=\textwidth]{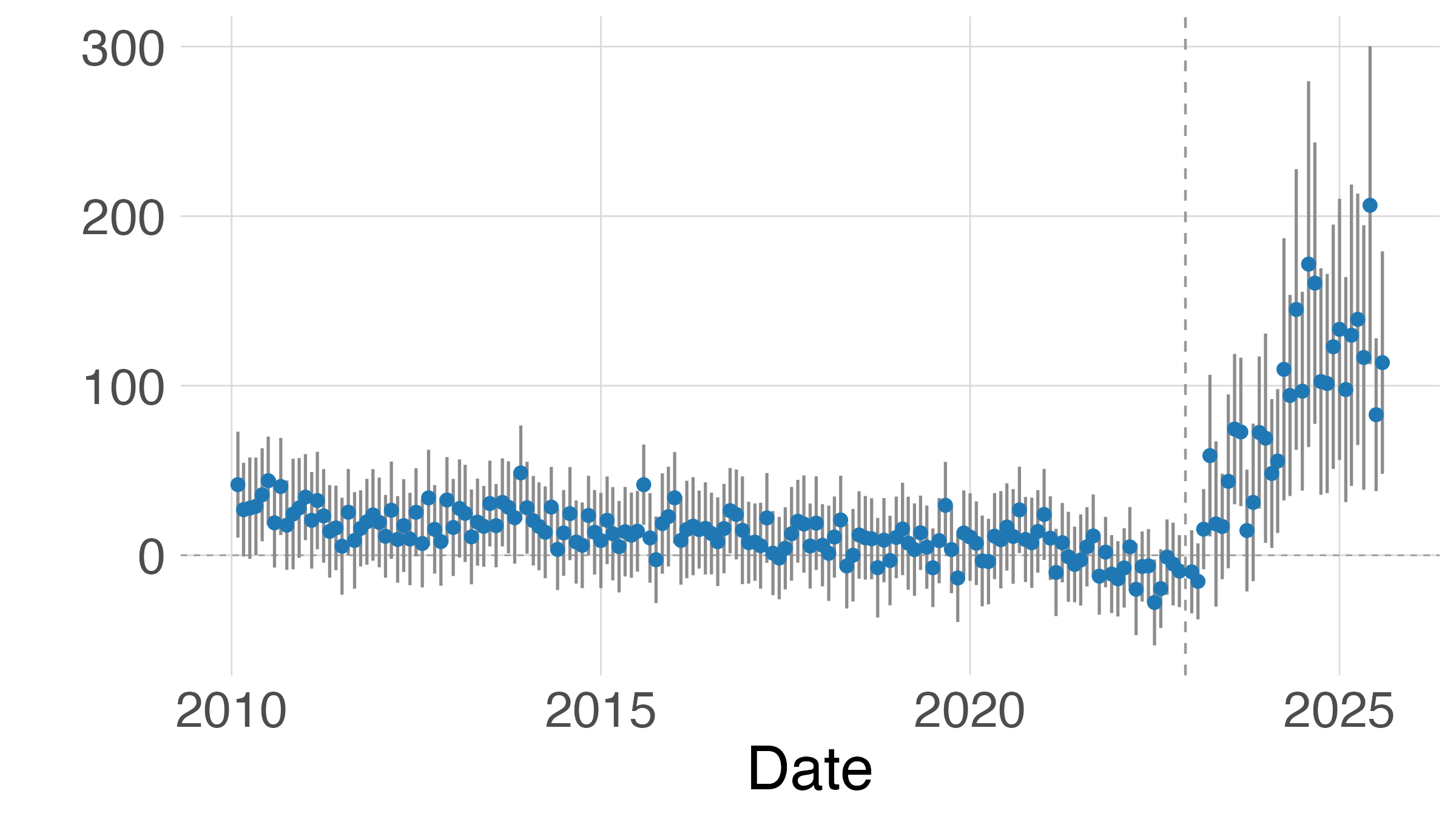}
        \caption{Event Study --- Earnings, Exposure to Automation}
        \label{fig:did_pretrends_earnings_aug}
    \end{minipage}
    \\
    \justifying
    \footnotesize\textit{ 
    Note: Figure \ref{fig:did_pretrends_earnings_aug}presents event-study estimates of unemployment rates for occupations with high-exposure (above the median exposure measured for automation tasks) versus low-exposure, relative to the introduction of ChatGPT in November 2022 (normalized to 0 at one period before treatment). Figure \ref{fig:did_pretrends_earnings_aug} presents the same event-study specification for average weekly earnings. Coefficients are shown relative to the pre-treatment baseline, with separate lines for treated and control occupations. The gray lines indicates the 95\% confidence interval, computed using clustered standard errors.
}
\end{figure}

\begin{table}[H]
\centering
\footnotesize
\begin{tabular}{lcc}
\toprule
& \textbf{Weekly Earnings} & \textbf{Unemployment} \\
\midrule
Above Median LLM Exposure $X$ Post    &  95.653 & 0.012 \\ & (19.949)  &  (0.002) \\
LLM Exposure $X$ Post & 288.890 & 0.035\\ & (65.117)  &  (0.005) \\
\midrule
Occupation Fixed Effect & V & V \\ 
YearXMonth Fixed Effect & V & V \\
\bottomrule
\end{tabular}
\caption{Estimates with standard errors in parentheses.}\label{tab:treatment_estimates}\justifying
    \footnotesize\textit{ 
    Note: The table shows the results from the naive DiD analysis. Clustered standard errors in parenthesis.}

\end{table}

\begin{figure}[H]
    \centering
    \begin{minipage}[b]{0.45\textwidth}
        \centering
        \includegraphics[width=\textwidth]{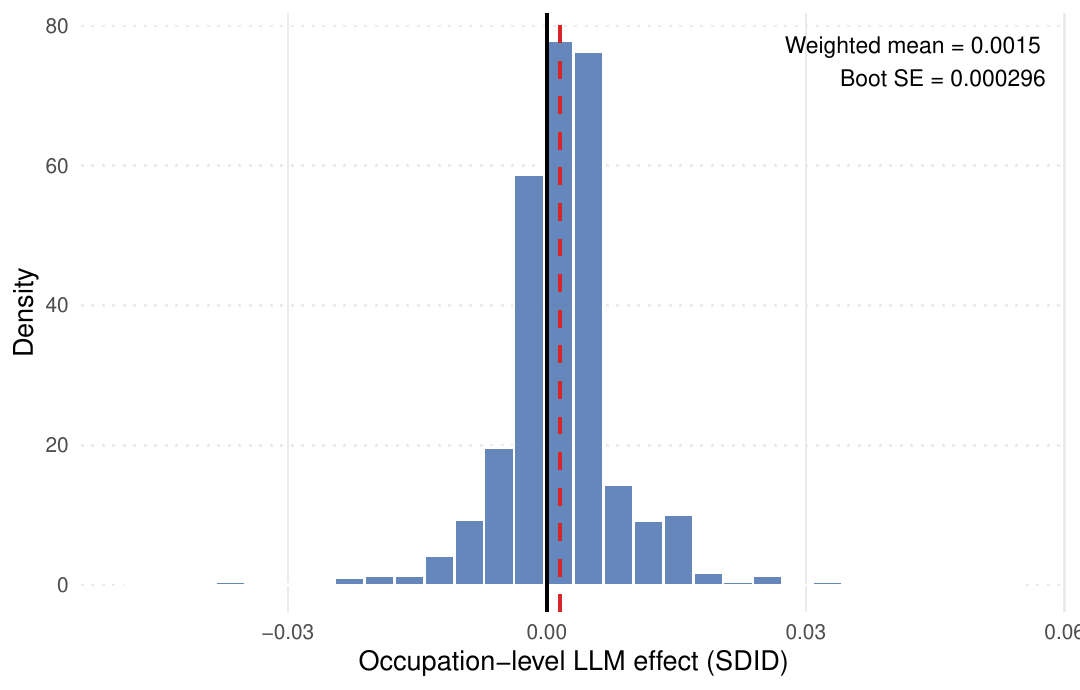}
        \caption{Distribution of SDiD Estimates: Unemployment, Augmentation Exposure}
        \label{fig:SDiD_unemployment_aug}
    \end{minipage}
    \hfill
    \begin{minipage}[b]{0.45\textwidth}
        \centering
        \includegraphics[width=\textwidth]{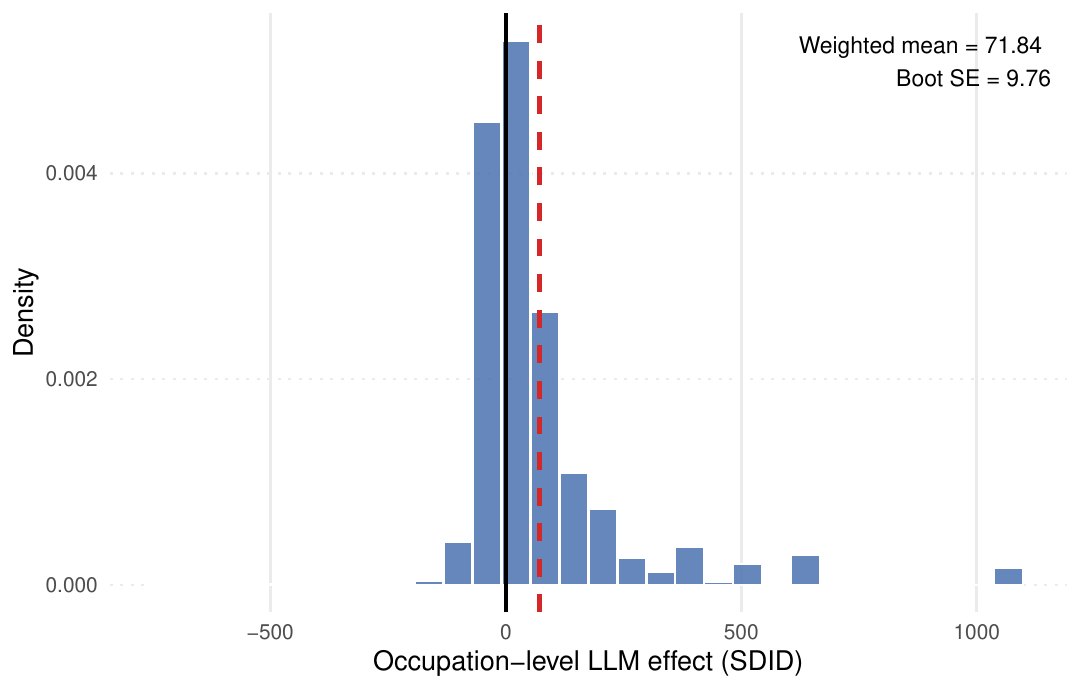}
        \caption{Distribution of SDiD Estimates: Weekly Earnings, Augmentation Exposure}
        \label{fig:SDiD_earnings_aug}
    \end{minipage}\\
    \justifying
    \footnotesize\textit{Note: Figure \ref{fig:SDiD_unemployment_aug} shows a histogram of Synthetic Difference-in-Differences (SDiD) estimates of the effect of LLM exposure on unemployment at the occupation level. Each bar represents the number of treated occupations with a given estimated effect. The horizontal axis reports the estimated change in the unemployment rate (percentage points), and the vertical axis shows the density across occupations. The red vertical line indicates the weighted average treatment effect on the treated (ATT). Figure \ref{fig:SDiD_earnings_aug} shows the corresponding histogram of SDiD estimates for weekly earnings at the occupation level. The horizontal axis reports the estimated change in real weekly earnings (January 2010 dollars), and the vertical axis shows the density of occupations. The red vertical line indicates the weighted average ATT, representing the mean earnings effect across treated occupations. The standard errors of the ATT are computed using a bootstrap procedure with 1,000 repetitions.}
\end{figure}

\begin{figure}[H]
    \centering
    \begin{minipage}[b]{0.45\textwidth}
        \centering
        \includegraphics[width=\textwidth]{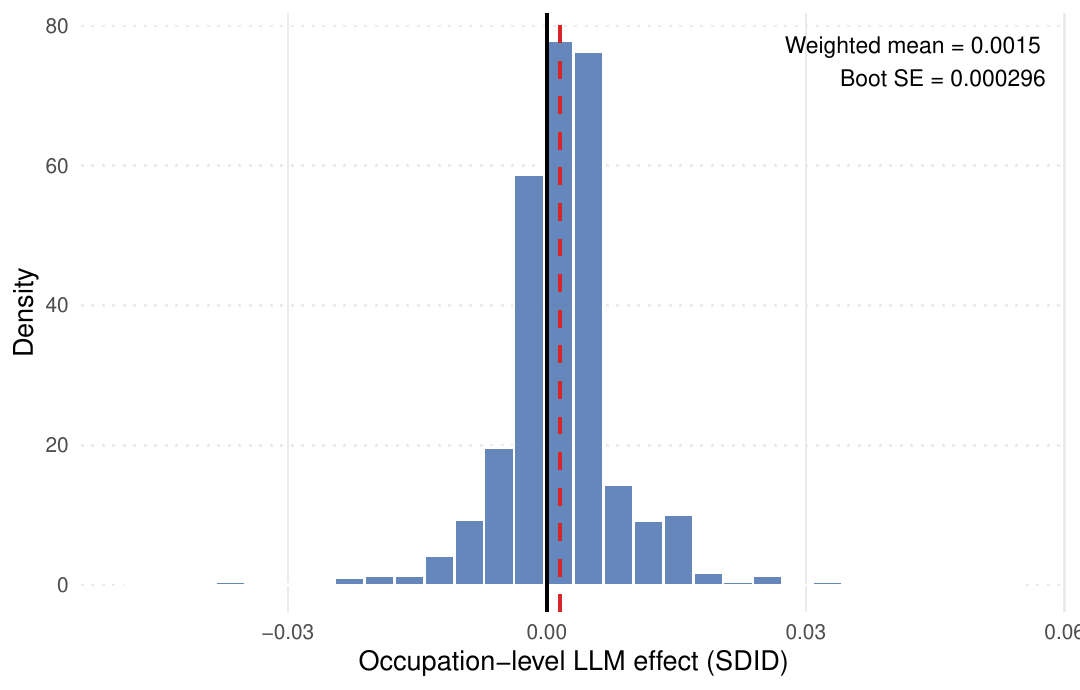}
        \caption{Distribution of SDiD Estimates: Unemployment, Automation Exposure}
        \label{fig:SDiD_unemployment_aut}
    \end{minipage}
    \hfill
    \begin{minipage}[b]{0.45\textwidth}
        \centering
        \includegraphics[width=\textwidth]{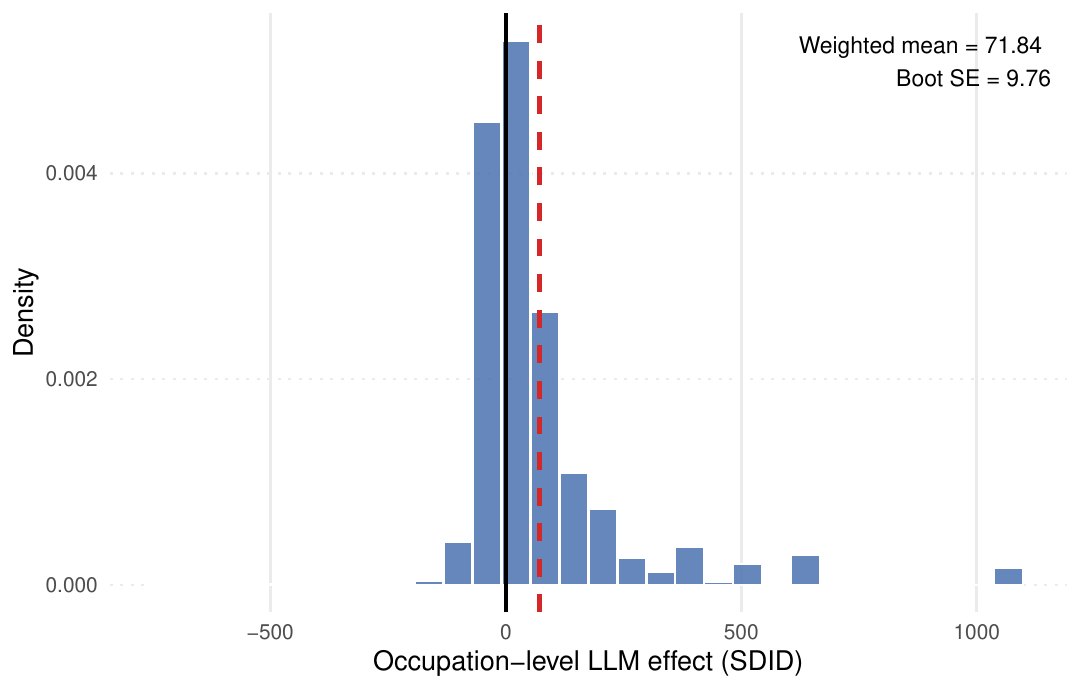}
        \caption{Distribution of SDiD Estimates: Weekly Earnings, Automation Exposure}
        \label{fig:SDiD_earnings_aut}
    \end{minipage}\\
    \justifying
    \footnotesize\textit{Note: Figure \ref{fig:SDiD_unemployment_aut} shows a histogram of Synthetic Difference-in-Differences (SDiD) estimates of the effect of LLM exposure on unemployment at the occupation level. Each bar represents the number of treated occupations with a given estimated effect. The horizontal axis reports the estimated change in the unemployment rate (percentage points), and the vertical axis shows the density across occupations. The red vertical line indicates the weighted average treatment effect on the treated (ATT). Figure \ref{fig:SDiD_earnings_aut} shows the corresponding histogram of SDiD estimates for weekly earnings at the occupation level. The horizontal axis reports the estimated change in real weekly earnings (January 2010 dollars), and the vertical axis shows the density of occupations. The red vertical line indicates the weighted average ATT, representing the mean earnings effect across treated occupations. The standard errors of the ATT are computed using a bootstrap procedure with 1,000 repetitions.}
\end{figure}

\bibliographystyle{plainnat} 
\bibliography{sample}    
\end{document}